\documentclass[%
preprint,
 amsmath,amssymb,
 aps,
]{revtex4-2}

\usepackage{graphicx}
\usepackage{dcolumn}
\usepackage{bm}

\begin{document}


\title{All-optical scanning vector magnetometry based on fine and hyperfine interactions in spin-$\frac{3}{2}$ centers in silicon carbide}

\author{Kirill V. Likhachev}
\email{kirilll28.1998@gmail.com}
\affiliation{Ioffe Institute, 194021 St. Peterburg, Russia}

\author{Maxim V. Uchaev}%
\affiliation{Ioffe Institute, 194021 St. Peterburg, Russia}

\author{Igor P. Veyshtort}
\affiliation{Ioffe Institute, 194021 St. Peterburg, Russia}

\author{Anastasia V. Batueva}
\affiliation{Ioffe Institute, 194021 St. Peterburg, Russia}

\author{Aleksandr S. Gurin}
\affiliation{Ioffe Institute, 194021 St. Peterburg, Russia}

\author{Roman A. Babunts}
\affiliation{Ioffe Institute, 194021 St. Peterburg, Russia}

\author{Pavel G. Baranov}
\affiliation{Ioffe Institute, 194021 St. Peterburg, Russia}


\begin{abstract}

The possibility of using axial spin color centers with $S=3/2$, oriented 
along the hexagonal $c$~axis in a silicon carbide (SiC) wafer, has been 
demonstrated for all-optical measurement of projection of the external 
magnetic field coinciding with the $c$~axis of the crystal, and the polar 
and azimuthal angles of the external measured magnetic field at room and 
significantly higher temperatures. A distinctive feature of spin centers 
in SiC, in which optically induced spin alignment is carried out, is the 
presence of a unique system of spin levels in a magnetic field, caused by 
fine structure interaction and hyperfine interaction with the $^{29}$Si 
nuclei and there is a wide range of level anticrossings (LACs), leading 
to an exceptionally strong change in photoluminescence in the region of 
LAC and the dependence of the LAC spectrum on the orientation of the 
external measured magnetic field, which made it possible to develop the 
all-optical vector magnetometry method. Such measurements do not require 
microwave radiation, it is possible to use single spin center for the 
all-optical vector magnetometry. The proposed magnetometer is based on an 
external magnetic field cancellation scheme, which leads to a LAC spectrum 
observed in a zero external magnetic field, called its reference spectrum, 
by maintaining a local region of zero magnetic field at the site of optical 
excitation of spin centers. 4H-SiC plate is placed on the scanning stage 
of a confocal microscope inside the Helmholtz coils. Sensitivity to a 
constant magnetic field for $z$ component of the magnetic field ${B}_{z}$ is 
better than $0.1\mu$T$/\sqrt{\text{Hz}}$ in a volume of $1\times10^{-8}$mm$^{3}$ 
at room temperature. The sensitivity of determining the angles polar and 
azimuthal is determined by the sensitivity to determining the perpendicular 
component of the magnetic field, which is better 
than $5~\mu$T$/\sqrt{\text{Hz}}$.
\end{abstract}

\maketitle


\section{Introduction}

Optically addressable atomic-scale spin color centers in the form of nitrogen vacancy 
(NV) defects in diamond~\cite{Grub97, PhysRevB.60.11503,Jele01} are 
the basis for creating magnetic field sensors with 
high sensitivity and submicron spatial resolution~\cite{Cher04, Bala08, Maze08, Tayl08}.
A unique opportunity was realized to record nuclear magnetic resonance (NMR) on protons
on the nanometer scale~\cite{Mami13, Stau13}, 
opening up the possibility of NMR imaging at the submicron level.

Magnetometry with sensors on NV centers having spin $S=1$ is based on the method of 
optical detection of magnetic resonance (ODMR), the position of resonances in 
which depends on the external magnetic field. When using sensors based on single 
NV defects, only one projection of the magnetic field on the axis of the 
NV center can be measured. When using ensembles of NV centers, vector magnetometry 
is possible, since NV centers in a diamond lattice are oriented in four 
$\langle$111$\rangle$ directions~\cite{Maer10, Stei10}.

There are a number of works where NV centers can be manufactured with preferential
alignment along one of the $\langle$111$\rangle$ axes~\cite{mich14, lesi14}, 
using ODMR line shift in relatively high magnetic fields (hundreds of Gauss) it is 
possible to reconstruct
the transverse component of the magnetic field~\cite{Teti12}. However, in basic applied 
developments it is necessary to obtain information about the vector parameters 
of small magnetic field values, which is difficult to achieve in the mentioned approach.

The use of axial spin color centers with high spin $S ={3}/{2}$ in silicon carbide of 
hexagonal and rhombic polytypes, first proposed in~\cite{simin2015high, kraus2014magnetic}, 
makes it possible to obtain magnetic field vector information based on the structure 
of spin levels in a magnetic field, while nature itself has aligned the axial axes
of these centers along the hexagonal $c$~axis of the crystal~\cite{baranov2007spin,
baranov2011silicon}.

In~\cite{simin2015high, kraus2014magnetic} it is shown that axial color centers with 
spin $S ={3}/{2}$ in silicon carbide with a hexagonal lattice structure can be used to 
measure not only the magnitude, but also the polar angle of the external magnetic field 
relative to the defect axis with high accuracy. The method is based on the optical 
detection of groups of magnetic resonances in a spin center based on a negatively 
charged silicon vacancy with a quartet ground spin state. The spin-$\frac{3}{2}$ color 
centers under study have axial symmetry along the $c$~axis of the crystal; these centers 
include a negatively charged silicon vacancy with spin $S ={3}/{2}$. The axial symmetry 
of the center is determined by the presence of excitation in the form of an intrinsic 
defect along the $c$~axis. This intrinsic defect, in our opinion, is a neutral carbon 
vacancy that is not covalently bonded to the aforementioned silicon 
vacancy~\cite{soltamov2015optically}.

An angular resolution of several degrees was achieved over a magnetic field range 
of up to several millitesla. This approach is applicable both for ensembles and for
single color centers with spin $S ={3}/{2}$. In~\cite{lee2015vector}, electronic 
spin systems with a high spin ($S >{3}/{2}$) are considered, which are promising 
for the development of vector magnetometry methods, since EPR spectra allow one 
to obtain information about both the magnitude of the magnetic field and its 
orientation. Analytical solutions are presented that make it possible to 
reconstruct the magnetic field strength and polar angle using the observed resonant
transitions in the $S =1$ and $S ={3}/{2}$ systems. The electron spins of a silicon
vacancy in silicon carbide with $S ={3}/{2}$ are presented as a vector magnetometer
model, the magnitude of the magnetic field can be extracted in the range of high 
and low magnetic field, the polar angle can only be determined in the low magnetic 
field, and the azimuthal angle cannot be determined in any case~\cite{lee2015vector}.
It is important to emphasize that only the use of microwave radiation in vector 
magnetometry was considered. As mentioned above, this method was practically 
implemented in our work~\cite{simin2015high, kraus2014magnetic} 
and in~\cite{niethammer2016vector}.

Previously, we proposed quartet spin centers in SiC to create all-optical methods for 
recording magnetic fields~\cite{anisimov2016spin, simin2016all} and temperatures 
\cite{anisimov2016optical} with micron and submicron spatial resolution. In 
\cite{anisimov2016spin, simin2016all}, a new approach was demonstrated in which the magnetic
field was measured in the absence of microwave power, i.e., completely optical 
magnetometry and thermometry under ambient conditions was demonstrated 
\cite{anisimov2016spin, simin2016all, anisimov2016optical, baranov2017magnetic}. However, 
only the longitudinal component of the magnetic field was measured, parallel to the axis of 
the spin centers, that is, parallel to the $c$~axis of the hexagonal 4H-SiC crystal, and the 
method does not provide information about the polar and azimuthal angles of the magnetic field.

The paper~\cite{baranov2017} demonstrated all-optical spectroscopy of hyperfine interactions 
under ambient conditions in SiC crystal with natural isotope content. All possible 
anti-crossings of spin levels in the ground state were recorded in the system of hyperfine 
interactions with one nucleus of the silicon isotope $^{29}$Si, which falls into the second 
coordination sphere relative to the silicon vacancy in the structure of the quartet 
spin center (twelve Si$_{NNN}$ atoms, where NNN is the next-nearest neighbor). The presented
studies open up opportunities for all-optical spectroscopy of hyperfine interactions in 
a wide range of quartet spin centers in other SiC polytypes, currently numbering more than 
ten discovered objects (in the 6H-, 15R-, 21R polytypes)~\cite{baranov2017magnetic}. In this 
work, based on all-optical spectroscopy of fine-structure and hyperfine interactions in silicon 
carbide~\cite{baranov2017} we have developed a new unique direction for creating sensors with
all-optical entirely vector magnetometry with micron and nanometer spatial resolution.

\section{Experimental}

The experiments presented here were carried out at room temperature on a radio spectroscopic 
complex created in Ioffe institute, in the form of an ODMR spectrometer, made on the basis 
of a scanning confocal optical microscope from NT MDT SI. The spin-$\frac{3}{2}$ color 
centers in the most common silicon carbide polytype, 4H-SiC, which has a hexagonal crystal 
lattice, has been used as the active material for fmeasurements. We used both crystals in 
which the wafer plane was perpendicular to the hexagonal $c$~axis, that is, parallel to the 
$z$~axis of the laboratory coordinate system, and commercial 4H-SiC crystals 
grown using standard sublimation techniques such that the crystallographic direction [0001] 
($c$~axis) was inclined at an angle of 4 degrees to the surface normal, that is, to 
the $z$~axis of the laboratory coordinate system. The axial axis of the 
spin-$\frac{3}{2}$ centers is oriented along the $c$~axis of the crystal. The scanning magnetic 
field ${B}_{0}$ and the modulating magnetic field are oriented parallel 
to the $c$~axis.

Figure~\ref{fig:Block diagram} shows a block diagram of the all-optical vector magnetometer. 
There are two modes for level anti-crossing (LAC) signal acquisition. In mode 1 the 
quasi-stationary magnetic field ${B}_{0}$ is modulated at low frequency (LF). 
In mode 2 the intensity of the exciting laser radiation is modulated at low frequency. 
For both options, the signal is recorded at low frequency using a lock-in detector. 4H-SiC 
wafer with spin-$\frac{3}{2}$ centers is used as the SiC sensor for measurements, crystal 
$c$~axis perpendicular to the wafer plane. The inset shows a structure in which a 4H-SiC 
wafer is used as the SiC sensor for measurements, in which the $c$~axis is deviated from 
the surface normal by 4 degrees (this configuration is often found 
in SiC device structures).

 The optical magnetometer includes an active material in the form of a silicon carbide 
 crystal wafer of a hexagonal or orthorhombic polytype containing spin-$\frac{3}{2}$ 
 centers and placed in an electromagnet that scans the a quasi-stationary magnetic field 
 ${B}_{0}$ along the $z$~axis, which is combined with 
 the electromagnet in the form of Helmholtz coils, creating a field along the 
 ${x, y, z}$~axes. These coils are designed to create an external 
 magnetic field when calibrating the magnetometer, as well as to create a compensating 
 magnetic field during measurements. A modulated magnetic field ${B}_{Mod}$ 
 can be applied to the coil for $z$~axis. As the low-frequency driving 
 current in the Helmholtz coils is directly proportional to the magnetic field it generates,
 current value will serve as an indirect measure of the magnetic field being cancelled 
 in each dimension.

The set-up includes laser emitting in the near-infrared region with $\lambda = 785$~nm, 
infrared transmission filter, the signal was collected using a 0.75 NA 100x lens, pinhole, 
photodetector (PD), made, for example, in the form of a photomultiplier, photodiode, avalanche 
photodiode was installed at the output of the spectrometer to record integral 
photoluminescence; low-frequency generator, lock-in detector to which a PD and a low-frequency 
generator were connected for amplitude modulation of the magnetic field, the relative change 
in PL at the resonant value of the magnetic field was recorded, control unit, direct current 
source. The optical magnetometer may also include a scanning stage of a confocal microscope 
with a piezoelement, capable of reciprocating movement in three mutually perpendicular 
directions under the influence of control voltages of the piezoelement, on which the active 
material is located. Modulation of the magnetic field is carried out by an electromagnet. 

The proposed magnetometer is based on an external magnetic field cancellation scheme, which
leads to an LAC spectrum observed in a zero external magnetic field, called its reference 
spectrum, by maintaining a local region of zero magnetic field at the site of optical 
excitation of spin centers. A sensor in the form of a 4H-SiC polytype silicon carbide 
plate containing spin centers is placed on the scanning stage of a confocal microscope from 
NT-MDT SI. The magnetic field for the three ${xyz}$ directions in the sensor 
is created by a set 
of Helmholtz coils. The created magnetic field is directly proportional to the current in 
the coils, so this current can serve as an indirect measure of the created magnetic field. 
A compensation current is passed through the coils to maintain a local region of zero 
magnetic field in the local area of optical excitation of spin centers.

Spin-$\frac{3}{2}$ color centers were introduced into a 4H-SiC single crystal with a low nitrogen 
concentration by irradiating the crystal with electrons with an energy of 2 MeV and a flux
of $\sim10^{18}$~cm$^{-2}$. It is customary to designate color centers with spin $S ={3}/{2}$ by 
the corresponding zero-phonon PL lines (zero-phonon lines – ZPLs); in our experiments, V2 centers
in 4H-SiC are studied.

\section{Results and analysis}

Figure~\ref{fig:Diagram of the energy levels} shows a diagram of the energy levels of the V2 spin
center in 4H-SiC, having $S ={3}/{2}$ in both the ground state (GS) and the excited state (ES). 
Non-resonant transitions from the ground state to the excited state occur upon excitation by an 
IR laser into the phonon subbands zone, followed by relaxation to lower levels of the $^{4}E$ excited
state. Reverse transitions from the excited state $^{4}E$ to the ground state $^{4}A$ are possible in two ways:
(i) radiative transition with spin conservation in the form of photoluminescence (PL), which is shown
in the inset to the figure, lifetime in the excited state is $\sim$6~ns; (ii) a non-radiative transition 
without spin conservation from the excited state $^{4}E$ to an intermediate metastable state, the so-called
inter-system crossing (ISC), followed by a transition from the metastable state to the ground state $^{4}A$.
In this case, due to the spin-selective nature of the transitions, after several cycles of optical excitation
(usually within the submicrosecond time range), optically induced spin alignment is carried out. 
Thus, for V2 in 4H-SiC, the lower levels with $M_{S}= \pm 1/2$ are populated, and the levels $M_{S}= \pm 3/2$
are emptied~\cite{baranov2017magnetic}.
Before optical excitation, the populations of spin levels at room temperature obeyed the Boltzmann distribution
and were practically identical, since the splitting of the fine structure between the levels $M_{S}= \pm 1/2$ and
$M_{S}= \pm 3/2$, equal to $\Delta=70$~MHz, is extremely small. 

Figure~\ref{fig:Diagram of the energy levels} presents energy levels for the ground state in the 
implementation of two structures: (i) in the second coordination sphere relative to a silicon vacancy 
of silicon atoms Si$_{NNN}$, there are no nuclei of the $^{29}$Si isotope, which has a magnetic moment; 
(ii) in the second coordination sphere, there is one nucleus $^{29}$Si ($I=1/2$) with a nuclear magnetic moment, 
a hyperfine interaction with the isotopic constant of $\sim$9~MHz is realized. To calculate energy levels and 
the corresponding wave functions, two software packages were used: VISUAL EPR~\cite{grachev1987correct} 
and EasySpin~\cite{stoll2006easyspin}. 

To align spin centers and excite photoluminescence (the PL spectrum is shown in the inset to 
Fig.~\ref{fig:Diagram of the energy levels}) 
a laser with a wavelength of 785 nm, a power of 150 mW, and a spot diameter in a scanning confocal 
microscope of $\sim$1 $\mu$m was used. 

Figure~\ref{fig:Dependence PL}(a) shows the experimental LAC spectrum for spin-$\frac{3}{2}$ center 
in a 4H-SiC crystal with a natural isotope content, recorded at room temperature in the orientation of 
a quasi-stationary magnetic field ${B}_{0}$ parallel to the c axis of the crystal, which
coincides with the direction of the axial axes of the spin-$\frac{3}{2}$ centers. Mode 1 is used, in 
which the magnetic field is modulated at a low frequency. The direction of the LF modulation magnetic 
field ${B}_{Mod}$ is coaxial with the direction of the quasi-stationary magnetic field
${B}_{0}$. The external magnetic field ${B}_{Ext}$, which is planned 
to be measured using the magnetometer in question, for the measurements presented in 
Fig.~\ref{fig:Dependence PL}(a) is equal to zero. Sharp changes in the PL intensity are observed in 
the region of level anti-crossings. The two upper spectra were recorded at a low temperature of 290~K
(blue line) and a high temperature of 350~K (red line) using a magnetic field modulation amplitude, 
less than 0.01~mT. Below is the LAC spectrum recorded at a temperature of 300~K (black line) and at 
the higher magnetic field modulation amplitude, $\sim$0.05~mT, which we take as the reference spectrum.

The EPR spectrum of spin centers with $S=3/2$ is described by the spin Hamiltonian: 

\begin{equation}
H=g\mu_B{\bf{S}}\cdot{\bf{B}}
+D[S^{2}_\text{Z}-1/3\cdot{S(S+1)}]+
A\bf {I}\cdot\bf{S}
\label{eq:first}
\end{equation}

where $\bf{B}$ is the magnetic field, g is the isotropic electronic g-factor equal to 2.00,
$\mu_B$ is the Bohr magneton, $\it{D}$ is the fine structure parameter in an axial 
crystal field, $\it{A}$ is the isotropic hyperfine interaction splitting with one nucleus 
of the $^{29}$Si isotope ($I=1/2$) located in the second coordination sphere of a silicon vacancy
included in the structure of the spin-$\frac{3}{2}$ center. The zero-field splitting (ZFS) for 
V2 centers in 4H-SiC $\Delta\nu=2D=23.4 \times10^{-4}$cm$^{-1}$~=~70~MHz. In accordance with the 
proposed structure of the spin centers with $S = 3/2$ in SiC, the main role is played by 
a negatively charged silicon vacancy $V_{Si}^{-}$, in the immediate environment of which there are 
four carbon atoms. We previously presented all-optical detection of hyperfine interactions 
in spin-$\frac{3}{2}$ centers in 6H-SiC crystals with one the nearest neighbor $^{13}$C nucleus,
C$_\text{NN}$ atom, in a 6H-SiC crystal tenfold enriched in the $^{13}$C isotope 
\cite{babunts2021fully}. The hyperfine structure for interaction with $^{29}$Si
in the second coordination sphere relative to a silicon vacancy, Si$_{NNN}$ atom, is
$\it{A}\cong$~9~MHz~=~$3\times10^{-4}$~cm$^{-1}$~\cite{baranov2017magnetic}.

System of energy levels calculated for spin-$\frac{3}{2}$ center in a 4H-SiC is shown in 
Fig.~\ref{fig:Dependence PL}(b). Solid lines correspond to the structure of the spin center, in which
there is hyperfine interaction with one $^{29}$Si nucleus, $I=1/2$; dotted lines correspond to 
a structure in which there is no hyperfine interaction, since there are no $^{29}$Si nuclei in 
the second coordination sphere of twelve silicon atoms, $I=0$. The gray circles show the level 
anti-crossings LAC1 and LAC2 for the ground state in the absence of hyperfine interaction. Red circles 
show level anti-crossings in the presence of one silicon atom with a nuclear magnetic moment of the 
$^{29}$Si isotope in the second coordination sphere. Two circles with lines of great thickness, leading 
to a doubling of the intensity of LACs in Fig.~\ref{fig:Dependence PL}(a) correspond to 
double level anti-crossings.

LAC spectroscopy of complex systems with high electron spin and hyperfine interactions with nuclear
spins opens up wide opportunities for obtaining information about the orientation of external magnetic
fields. Minor deviations in the orientation of the c-axis of the crystal, and, consequently, the axial
axis of the spin-$\frac{3}{2}$ center from the direction of the magnetic field, lead to significant 
changes in the LAC spectrum, which can be used for all-optical vector magnetometry, that is, in the 
complete absence of microwave radiation.

This type of vector magnetometry is based on the principle of compensating for the external magnetic 
field that need to be measured using a system of calibrated Helmholtz coils that create a magnetic 
field in three $xyz$ directions. As a result of compensation of the external measured 
magnetic field, it is expected to obtain an accurate level anti-crossing system, which we take as 
the reference spectrum (Fig.~\ref{fig:Dependence PL}(a)).

Figure~\ref{fig:Orientation dependences} shows the orientation dependences of the LAC spectra 
when the crystal rotates in a 
quasi-stationary magnetic field ${B}_{0}$, as shown in the inset. The direction of the 
low-frequency modulation magnetic field ${B}_{Mod}$ coincides with ${B}_{0}$. 
A strong dependence of the shape of the LAC signals on the orientation of the crystal in the magnetic
field is observed. This dependence is explained by the structure of Zeeman levels for 
spin-$\frac{3}{2}$ centers in the 4H-SiC crystal.

Figure~\ref{fig:energy levels calc} shows the energy levels for spin-$\frac{3}{2}$ 
center in a 4H-SiC calculated for 
different crystal orientations in a magnetic field. Solid lines correspond to the structure 
of the spin center, in which there is hyperfine interaction with one $^{29}$Si nucleus 
(Si$_{NNN}$), $I=1/2$; dotted lines correspond to a structure in which there is no hyperfine
interaction, since there are no $^{29}$Si nuclei in the second coordination sphere of twelve 
silicon atoms, $I=0$. The black dotted lines show the corresponding LAC spectra taken from 
Fig.~\ref{fig:Orientation dependences}. The proposed magnetometer is based on a magnetic 
field cancellation scheme by 
maintaining a local region of zero-magnetic field at the point of focusing the laser excitation 
of spin-$\frac{3}{2}$ centers in a magnetic field sensor, which leads to the reference LAC spectrum 
presented in Fig.~\ref{fig:Dependence PL}(a).

Figures~\ref{fig:principles of calibrating} and~\ref{fig:Calibrating magnetometer} demonstrate 
the principles of calibrating a vector magnetometer using a system of Helmholtz magnetic coils 
to create a given magnetic field in three  ${x, y, z}$~directions. For ease of measurement,
the calibrated magnetic field will be applied in the $z$ and ${x}$~directions, 
resulting in a change in the polar angle $\theta$. Since the measurements are based on axial spin 
centers whose axis coincides with the crystal $c$~axis, the measurement results depend only on the 
ratio of the longitudinal and perpendicular components of the external magnetic field, that is, 
they depend on the polar angle $\theta$ and do not depend on the azimuthal angle $\varphi$.

When carrying out the measurements in Fig.~\ref{fig:Calibrating magnetometer}(a), the 
${B}_{z}$-component of the magnetic field ${B}_{z}$ is reset to zero by shifting the LAC1 
point to the standard value $D$ (1.25 mT). Next, only the perpendicular component changes. The final polar
angle $\theta$ is determined each time from the found ${B}_{z}$ and ${B}_{\perp}$. 
The appearance of the LAC spectrum depends only on the polar angle $\theta$. The azimuthal angle $\varphi$
is found by compensating the perpendicular component of the external magnetic field by rotating the 
compensating perpendicular component in the azimuthal plane. When the axis of the perpendicular component 
of the external field coincides with the axis of the compensating magnetic component with their opposite 
directions, but equal absolute values of these components, the external magnetic field is zeroed and the 
LAC spectrum has a form that coincides with the reference LAC spectrum (Fig.~\ref{fig:Dependence PL}(a)).
Reference LAC spectrum is measured 
in the absence of an external magnetic field and the orientation of the quasi-stationary magnetic field
${B}_{0}$ (scanning field) parallel to the axial axis of the spin center, coinciding with 
the crystalline $c$~axis of the SiC based sensor.

Figure~\ref{fig:Calibrating magnetometer}(b) demonstrates a changes in the intensities $I$ of selected 
lines in the LAC spectra (3, 5, 6, 7 in the top of Fig.~\ref{fig:Calibrating magnetometer}(a)), 
characterizing changes in the photoluminescence intensity of spin centers with a ground quadruplet 
spin-$\frac{3}{2}$ state, depending on perpendicular magnetic field ${B}_{x}$ (mT), which serve 
as criteria for the preliminary assessment of external measured magnetic fields and the selection of the 
values of compensating magnetic fields ${B}_{Comp}$ for zeroing the measured external magnetic field.

$Criterion~\it1.$ Line intensity (5) in Fig.~\ref{fig:Calibrating magnetometer}(a), designated 
as $I5$ in Fig.~\ref{fig:Calibrating magnetometer}(b), decreases (8), as the magnitude of the perpendicular
magnetic field increases and with the perpendicular component ${B}_{x}$ equal to 0.15 mT it becomes
equal to zero; in perpendicular fields greater than 0.15 mT this line does not appear. Dependencies 
in~\ref{fig:Calibrating magnetometer}(b) are determined by the system of energy levels and are fixed for 
the used sensor based on a 4H-SiC crystal with a natural content of isotopes with spin-$\frac{3}{2}$ centers.
Based on this fact, we can formulate the first criterion: after introducing an external (measured) magnetic 
field, if we observe line (5) in the LAC spectrum - the perpendicular component of the external magnetic 
field is $<0.15$~mT. Thus, in the region of low fields (up to 0.15 mT), it is possible to carry out completely 
optical vector magnetometry based on line intensity (5). Considering that line (5) changes relatively 
little near the point ${B}_{x}=0$, it is convenient to carry out more accurate measurements by 
changing the magnetic field ${B}_{x}$ with a transition through the zero value, that is, within
$-0.1\text{~mT}<{B}_{x}<+0.1\text{~mT}$, that is, in magnetic fields, where the intensity of line (5) changes most 
sharply, followed by finding the midpoint, which will correspond to ${B}_{x}=0$.

$Criterion~\it2.$ The intensity of line (6) is compared with the intensity of line (3) as the 
perpendicular component of the calibration magnetic field increases in
Fig.~\ref{fig:Calibrating magnetometer}(a). We plot the dependence of the intensity ratio of these lines 
as ($I6$)/($I3$) on the value of the perpendicular component of the external magnetic field in 
Fig.~\ref{fig:Calibrating magnetometer}(b); with a calibration magnetic field ${B}_{x}=0.28$~mT, the 
intensity ratio becomes equal to~1. This leads to the second criterion: after introducing 
the perpendicular component of the external magnetic field, if the intensities of peaks (6) and (3) coincide, 
we are in the region of the perpendicular component of the magnetic fields ${B}_{x}=0.28$~mT, if 
the ratio is greater than 1, ${B}_{x}<0.28$~mT, if the ratio is less than 1, ${B}_{x}>0.28$~mT,
and then a monotonic decrease in the ratio of the intensities of lines (6) and (3) is observed with 
an increase in the perpendicular component of the magnetic field, which makes it possible to judge the 
magnitude of the perpendicular component of the magnetic field in a wide range of magnetic fields. 

$Criterion~\it3.$ Line (7) in Fig.~\ref{fig:Calibrating magnetometer}(a) decreases as the magnitude 
of the perpendicular magnetic field, ${B}_{x}$, increases. The dependence of the ratio of the 
intensities of lines (6) and (7) on the magnitude of the perpendicular magnetic field in the form of the 
ratio ($I6$)/($I7$) is plotted in Fig.~\ref{fig:Calibrating magnetometer}(b), it is clear that at 
${B}_{x}=0.4$~mT line (7) merges with line (6) and with a further increase in ${B}_{x}$ is not
resolved. Thus, if line (7) is not resolved, the perpendicular component of the magnetic field is greater 
than or equal to 0.4 mT, that is, when ${B}_{x}>0.4$~mT, the splitting of the hyperfine structure for
the second LAC completely disappears, that is, the resolved structure for lines (6) and (7).

$Criterion~\it4.$ At ${B}_{x}=0.6$~mT, the splitting of the hyperfine structure for the first 
LAC completely disappears, that is, if the allowed structure for lines (2), (3) and (4) 
(Fig.~\ref{fig:Dependence PL}(a)) disappears, then the perpendicular component of the external magnetic 
field ${B}_{x}>0.6$~mT.

The next stage of measuring the external magnetic field is the sequential compensation of all three 
components ${B}_{z}$, ${B}_{x}$ and ${B}_{y}$, created by the external magnetic 
field ${B}_{Ext}$ of a permanent magnet located at two arbitrary points (Examples 1 and 2). 
The first stage of compensation of the external magnetic field is carried out by applying a compensating 
magnetic field that zeroes the parallel component ${B}_{z}$ of the external magnetic field, 
that is, the value of the compensating field along the $z$~axis changes until the points LAC1 
of the distorted spectrum and the reference spectrum coincide. The second stage of external magnetic 
field compensation is to introduce a compensating magnetic field in the perpendicular ${xy}$ 
plane, which changes the shape of the LAC spectrum without changing the position of the LAC1 in 
a quasi-stationary magnetic field, which was established by the compensating magnetic field component 
along the $z$~axis. Thus, by changing the amount of current supplied to the Helmholtz coils 
in the ${xy}$ plane, it is possible to achieve complete compensation of the external (measured) 
magnetic field, that is, its zeroing in the region excited by focused laser radiation. Full 
compensation occurs when the LAC spectrum, distorted by an external magnetic field, as a result of the 
introduction of compensation fields, completely coincides in shape and position with the reference 
LAC spectrum, recorded in a zero external magnetic field (Fig.~\ref{fig:Dependence PL}(a)).

Figure~\ref{fig:Demon princ} demonstrates the principles of all-optical quantum vector magnetometry. Two experiments 
($\#$1 and $\#$2) are presented to compensate for the external random created magnetic field by two positions 
of a permanent magnet. Top in Fig.~\ref{fig:Demon princ}(a) shows the spectrum of 
LAC (1) recorded in the absence of an external magnetic field (solid lines for all spectra corresponding 
to experiment $\#$1, dotted lines to experiment $\#$2), also shown in Fig.~\ref{fig:Dependence PL}, where 
the energy levels leading to mentioned LAC were demonstrated. 

The following spectra (2) were recorded with the application of an arbitrary external measured magnetic 
fields ${B}_{Meas_{1}}$ and ${B}_{Meas_{2}}$, conventionally shown in Fig.~\ref{fig:Demon princ}(b). 
The two lower spectra correspond to the sequential application of compensating magnetic fields in the form 
of the ${B}_{z_{1}}$ and ${B}_{z_{2}}$ component (3), shifting the signals of 
the first LAC1 until they coincide with the initial
reference signal (1), and then the ${B}_{x_{1}}$ and ${B}_{y_{1}}$ (${B}_{x_{2}}$ and 
${B}_{y_{2}}$) components (4) are applied until a spectrum 
appears that almost coincide with the spectra (1) , shown with a dotted line. In Fig.~\ref{fig:Demon princ}(b)
the principle of compensating a random measured magnetic field ${B}_{Meas}$ (red arrows) by applying a 
compensation field ${B}_{Comp}$ (blue arrows) is shown conventionally. 

The experiments were performed at room temperature, the area of the optical excitation spot was 
$\sim$1~$\mu$m$^2$, the magnetic field modulation frequency ${f}_{Mod}=331$~Hz, 
the modulation amplitude of 0.01~mT,
the modulation magnetic field was directed along the quasi-stationary magnetic field ${B}_{0}$, that is,
perpendicular to the plane of the SiC wafer, the normal of which coincides with the crystal $c$~axis. 
The appearance of the LAC spectrum depends only on the polar angle $\theta$ and does not depend on 
the azimuthal angle $\varphi$. The azimuthal angle is found by compensating the perpendicular component of 
the external magnetic field by rotating the compensating perpendicular component in the azimuthal plane. 
When the axis of the perpendicular component of the external field coincides with the axis of the compensating 
magnetic component with their opposite directions, but equal absolute values of these components, the external 
measured magnetic field (${B}_{Ext}$) is canceled and the LAC spectrum has a form that coincides with 
the reference spectrum in the absence of an external magnetic field and the orientation of the quasi-stationary 
magnetic field ${B}_{0}$ (scanning) parallel to the axial axis of the spin center, coinciding with 
the crystalline $c$~axis of the SiC-based sensor.

$Example~\#\it1.$ Because the magnetometer uses a magnetic field modulation circuit, the sensitivity of the 
instrument can be determined by the zero-crossing slope of the response and the noise remaining in the frequency 
band of interest of the modulation frequency used (fig.~\ref{fig:Demon princ}). 
As a result, by registering the ${B}_{z}$ component, 
a sensitivity to the magnetic field of $\sim0.1~\mu\text{T}/\sqrt{\text{Hz}}$ is achieved in a volume of 
$\sim1\times10^{-8}~\text{mm}^{3}$ at room temperature.
The module and direction of the compensating
field in the form of polar, $\theta$, and azimuthal, $\varphi$, angles are calculated using the formulas: 
$|{B}|=\sqrt{({B}_{x}^{2}+{B}_{y}^{2}+{B}_{z}^{2})}$, 
${B}_{\perp}=\sqrt{({B}_{x}^{2}+{B}_{y}^{2})}$, $\theta=\arctan{({B}_{\perp}/{B}_{z})}$, 
$\varphi=\arctan{({B}_{x}/{B}_{y}})$, ${B}_{z}=0.131$~mT, ${B}_{x}=0.174$~mT, ${B}_{y}=0.200$~mT,
$|{B}|=0.296$~mT; $\theta={63.70}^{\circ}$, $\varphi={49.00}^{\circ}$. 
The sensitivity of determining the angles polar $\theta$
and azimuthal $\varphi$ is determined by the sensitivity to determining the perpendicular component of the 
magnetic field, which is $\sim5~\mu\text{T}/\sqrt{\text{Hz}}$.

$Example~\#\it2.$ Measurement at a second arbitrary point near the magnetic field sensor.
${B}_{z}=0.110$~mT, ${B}_{x}=0.174$~mT, ${B}_{y}=0.207$~mT, $|{B}|=0.290$~mT; 
$\theta={67.90}^{\circ}$, $\varphi={50.00}^{\circ}$.

V2 spin-$\frac{3}{2}$ centers in a 4H-SiC crystal are not the only system for the proposed all-optical 
vector magnetometry scheme based on LAC spectroscopy. Spin color centers with $S = 3/2$, active under ambient 
conditions, were observed by us in most of the used hexagonal and orthorhombic SiC 
polytypes~\cite{baranov2017magnetic}. In particular, LACs were observed in the 6H-SiC crystal for three 
types of centers, V2, V1 and V3. The fine and hyperfine structure parameters for V1 and V3 are close in 
value~\cite{baranov2017magnetic, soltamov2021electron}, so we will further combine them under the general
designation V1/V3. It is for this polytype that we studied in detail the LAC spectra of spin-$\frac{3}{2}$ centers
with changes in the isotopic composition of $^{13}$C and $^{29}$Si isotopes with nuclear magnetic moments. 
Thus, direct experimental information was obtained on the effect of hyperfine interactions on the shape of 
the LAC spectra, which we propose to use for all-optical magnetometry on spin-$\frac{3}{2}$ centers.

Figure~\ref{fig:LAC recorded} shows (a) level anticrossing spectra 
recorded in the ground state for V2 centers in a 6H-SiC 
crystal, in the $B\parallel{c}$ orientation, excitation 808~nm, grown with a depleted 
$^{29}$Si isotope content, $\sim1\%$,
natural $^{13}$C isotope content, $1.1\%$ (red color) and in a 6H-SiC crystal grown with an enriched 
content of the $^{13}$C isotope up to $\sim15\%$, the natural content of the $^{29}$Si isotope is $4.7\%$ 
(blue color). Open circles and squares indicate anticrossings of LAC1 and LAC2 levels, respectively; 
circles and squares of reduced size show signals of level anti-crossing for V1/V3 centers. 
(b) Scheme of energy levels in the magnetic field of V2 centers taking into account hyperfine interactions
with the one nuclear $^{13}$C isotope ($I = 1/2$), level anti-crossing signals are marked by open diamonds. The dotted
line in red shows energy levels in a magnetic field without taking into account hyperfine interactions, 
which is typical for even silicon isotopes that do not have a nuclear magnetic moment, that is, there is no 
hyperfine interaction. It can be seen that a decrease in the content of the $^{29}$Si isotope, which has 
a nuclear magnetic moment, significantly narrows the line.

It is also important to emphasize the gigantic increase in line widths in a crystal enriched with the $^{13}$C
isotope. Thus, it is possible to reduce the width of the LAC line and thereby increase the accuracy of 
measuring magnetic fields by changing the isotopic composition of SiC, in particular, reducing 
the content of the $^{13}$C isotope.

The proposed magnetometer contains a set of capabilities for self-calibration, due to the known natural
interaction constants of the fine and hyperfine structure for the SiC sensor. The calibration is an important
problem when using magnetometers, this is especially important if measurements are made on remote scientific 
instruments, for example, in outer space or in hard-to-reach areas of the Earth. 
Figure~\ref{fig:Dependence PL}(a) illustrates these interactions, point LAC1 is characterized by a magnetic 
field equal to the parameter $D$, expressed in magnetic field units, LAC2 is characterized by a magnetic field 
equal to the parameter $2D$, that is, the distance between these points is equal to $D$. You can also use the 
distance between the two satellites from the center line LAC1, which is determined by the hyperfine interaction 
with one $^{29}$Si nucleus of 9~MHz. The first and simpler method of self-calibration involves the use of 
electron spin interactions, which are observed in the ODMR spectrum. Since all of these interactions are 
virtually independent of temperature, and are also extremely stable over long periods of time, the distances 
between them will be maintained with exceptional stability. By measuring the distance between these magnetic 
field indicators as a function of the applied coil current, a measure of the proportionality constant 
can be obtained that can be used to calibrate the coil drive system. An additional calibration method 
can be implemented by incorporating a miniature microwave excitation source, i.e. the standard ODMR method. 
This undoubtedly complicates the device, but the additional magnetic dimension can be extremely 
important in conditions, for example, in outer space.

The work~\cite{baranov2017} demonstrated the possibility of recording LAC signal under pulsed optical 
excitation, which results in a dynamic process of alignment of spin level populations and destruction 
of alignment in the LAC region within a short time of an optical pulse in the sub-millisecond range. 
This technique is expected to be used in the future to study transient spin processes at various 
points of the LAC, which is also a promising direction for use in vector magnetometry.

Unlike NV centers in diamond, where magnetic field and temperature measurements are made using 
properties of the ground spin-1 state, in silicon carbide it is possible to separate the local 
magnetic field and local temperature measurements using spin-$\frac{3}{2}$ centers: 
the magnetic field is measured by LAC in the ground state, and the temperature is measured 
by LAC in the excited state. The EPR spectra in the excited state are described in 
the same way as in the ground state by the spin Hamiltonian (1) with $S=3/2$. In this case, 
the temperature-dependent parameter $D(T)$ in the excited state is more than an order 
of magnitude greater than the parameter $D$ in the ground state (independent of temperature),
respectively, $D(T) =\sim15~\text{mT}$ (at RT) and $D = 1.25~\text{mT}$.

Under the conditions of recording the reference spectrum from LAC signals in the ground state 
while measuring the external magnetic field by compensating the latter, simultaneous measurements 
of the local temperature at the point of magnetic field measurement were carried out by the 
shift of the LAC signal~\cite{anisimov2016optical} in the excited state of 
spin-$\frac{3}{2}$ centers. 

Figure~\ref{fig:Temperature}~(a) shows the effect of temperature on the position of LAC in the ground 
and excited states in a 4H-SiC crystal. Shown is a ground state reference spectrum recorded at two 
temperatures, room temperature 300~K and elevated temperature 350~K, and an excited state LAC 
spectrum recorded at the same temperatures. It can be seen that an increase in temperature 
does not lead to a shift of LAC in the magnetic field, while for the excited state 
a significant shift is observed, reflecting the dependence of the parameter $D$ on temperature. 
A sensitivity of $\sim0.05~\text{mT/K}$ was achieved in 
the temperature range of 20-90$^{\circ}$C. Figure~\ref{fig:Temperature}~(b) demonstrates 
the temperature color map obtained by converting the position of LAC excited states 
into a temperature value from 
a sample area of $100\times100~\mu\text{m}$ with 
a step of $5~\mu\text{m}$. The measured contrast of 1$^{\circ}$C was achieved due 
to local heating and heat removal from the crystal through the Peltier element.

\section{Conclusion}

Near the level anticrossing, a sharp change in the intensity of photoluminescence is observed, which 
is used for all-optical sensing of the magnetic field. Sensitivity to a constant magnetic field 
(in this case, only one component of the magnetic field can be measured) better than $100~$nT/$\sqrt{\text{Hz}}$
in a volume of $\sim1\times10^{-8}~\text{mm}^{3}$ at room temperature has been achieved, while 
measurement stability has been demonstrated at high temperatures up to 
500~K~\cite{anisimov2016spin,simin2016all}. The contribution from hyperfine interactions 
with $^{29}$Si leads to an additional set of level anticrossing points with changes in both the 
electron spin projection and the projections of the nuclear spins of $^{29}$Si. At these LAC points,
a sharp change in photoluminescence intensity also occurs, and the magnitude of this change depends
on the isotopic enrichment of the silicon carbide. In silicon carbide 4H-SiC with a natural content 
of isotopes, there is a hyperfine interaction with one $^{29}$Si nucleus, which statistically falls 
into one of the twelve silicon positions in the second coordination sphere relative to the 
negatively charged silicon vacancy, which forms the basis of the axial 
spin-$\frac{3}{2}$ center structure~\cite{soltamov2015optically}.

The sharp dependence of the LAC spectrum on the orientation of the external magnetic field allows
to develop all-optical vector magnetometry without the use of microwave power. That is, we exclude
an increase in ODMR lines width due to the absorption of microwave power, which, as a rule, requires
the use of low microwave power to excite the signal, which leads to a decrease in the ODMR contrast.
There is no need to manufacture a microwave antenna, which greatly simplifies the design of 
the magnetometer, and also allows the use of large crystals. The compensation methodology 
for measuring external magnetic fields is used, which allows you to measure not only the polar 
angle $\theta$ , but also the azimuth angle $\varphi$. The sensitivity of determining the angles polar $\theta$ 
and azimuthal $\varphi$ is determined by the sensitivity to determining the perpendicular 
component of the magnetic field (${B}_{x}$,${B}_{y}$), which is better
than~$\sim5\mu\text{T}/\sqrt{\text{Hz}}$.

A vector magnetometer that simultaneously measures all Cartesian components of a magnetic 
field using ODMR on an ensemble of NV centers in a single diamond crystal was 
demonstrated~\cite{schloss2018simultaneous}. An article was recently 
published~\cite{garsi2024three}, in which sensors based on ODMR of NV centers in diamond were 
used to obtain a three-dimensional image of integrated-circuit activity. Obtaining information 
the distribution of currents when scaling semiconductor technologies to micron and submicron sizes is
of decisive importance for the development of modern microelectronics. In~\cite{garsi2024three},
the entire three-dimensional pattern of magnetic field distribution was projected onto the diamond 
plane with NV centers. It should be noted that obtaining information about the vector magnetic field
requires the use of an ensemble of NV centers and is not applicable for single centers. We believe
that our vector magnetometry technique based on spin-$\frac{3}{2}$ centers in silicon carbide can 
also be used to solve similar problems, without using microwave power. 
Magnetic field sensors based on spin centers in silicon 
carbide, measuring Oersted fields created by currents flowing inside multilayer integrated circuits 
in operating mode or with certain disturbances in operating mode, make it possible to build a 
three-dimensional picture of current densities. This is a sensitive, non-invasive method for
monitoring and diagnostics in the development and use of device structures. At the same time, 
it may be very promising to use this technique directly in chips and other device structures based 
on silicon carbide, into which spin centers are introduced by additional irradiation, or formed during
the manufacture of these structures. As a result, one can measure the vector magnetic field in 
specified areas excited by focused laser radiation, both at the stage of 
development of the device and in the operating device. It should be emphasized that obtaining 
information about the vector magnetic field does not require the use of an ensemble of spin centers
and can be applied to single spin-$\frac{3}{2}$ 
centers~\cite{widmann2015coherent, baranov2017magnetic}.

It is important to note that the proposed method for determining vector magnetic fields can also 
be applied to SiC nanocrystals that have at least one spin-$\frac{3}{2}$ center. It will first 
be necessary to establish the orientation of the SiC nanocrystal, and hence the orientation of 
the spin-$\frac{3}{2}$ center, by applying magnetic fields using Helmholtz coils until the reference
spectrum is obtained. In this case, the calibration dependences obtained in this work will be used.

\begin{acknowledgments}
This work was supported by the Russian Science Foundation № 23-12-00152 \\
(https://rscf.ru/project/23-12-00152/)
\end{acknowledgments}

\bibliography{apssamp}

\providecommand{\noopsort}[1]{}\providecommand{\singleletter}[1]{#1}%
\begin{thebibliography}{33}%
\makeatletter
\providecommand \@ifxundefined [1]{%
 \@ifx{#1\undefined}
}%
\providecommand \@ifnum [1]{%
 \ifnum #1\expandafter \@firstoftwo
 \else \expandafter \@secondoftwo
 \fi
}%
\providecommand \@ifx [1]{%
 \ifx #1\expandafter \@firstoftwo
 \else \expandafter \@secondoftwo
 \fi
}%
\providecommand \natexlab [1]{#1}%
\providecommand \enquote  [1]{``#1''}%
\providecommand \bibnamefont  [1]{#1}%
\providecommand \bibfnamefont [1]{#1}%
\providecommand \citenamefont [1]{#1}%
\providecommand \href@noop [0]{\@secondoftwo}%
\providecommand \href [0]{\begingroup \@sanitize@url \@href}%
\providecommand \@href[1]{\@@startlink{#1}\@@href}%
\providecommand \@@href[1]{\endgroup#1\@@endlink}%
\providecommand \@sanitize@url [0]{\catcode `\\12\catcode `\$12\catcode `\&12\catcode `\#12\catcode `\^12\catcode `\_12\catcode `\%12\relax}%
\providecommand \@@startlink[1]{}%
\providecommand \@@endlink[0]{}%
\providecommand \url  [0]{\begingroup\@sanitize@url \@url }%
\providecommand \@url [1]{\endgroup\@href {#1}{\urlprefix }}%
\providecommand \urlprefix  [0]{URL }%
\providecommand \Eprint [0]{\href }%
\providecommand \doibase [0]{https://doi.org/}%
\providecommand \selectlanguage [0]{\@gobble}%
\providecommand \bibinfo  [0]{\@secondoftwo}%
\providecommand \bibfield  [0]{\@secondoftwo}%
\providecommand \translation [1]{[#1]}%
\providecommand \BibitemOpen [0]{}%
\providecommand \bibitemStop [0]{}%
\providecommand \bibitemNoStop [0]{.\EOS\space}%
\providecommand \EOS [0]{\spacefactor3000\relax}%
\providecommand \BibitemShut  [1]{\csname bibitem#1\endcsname}%
\let\auto@bib@innerbib\@empty
\bibitem [{\citenamefont {Gruber}\ \emph {et~al.}(1997)\citenamefont {Gruber}, \citenamefont {Dräbenstedt}, \citenamefont {Tietz}, \citenamefont {Fleury}, \citenamefont {Wrachtrup},\ and\ \citenamefont {von Borczyskowski}}]{Grub97}%
  \BibitemOpen
  \bibfield  {author} {\bibinfo {author} {\bibfnamefont {A.}~\bibnamefont {Gruber}}, \bibinfo {author} {\bibfnamefont {A.}~\bibnamefont {Dräbenstedt}}, \bibinfo {author} {\bibfnamefont {C.}~\bibnamefont {Tietz}}, \bibinfo {author} {\bibfnamefont {L.}~\bibnamefont {Fleury}}, \bibinfo {author} {\bibfnamefont {J.}~\bibnamefont {Wrachtrup}},\ and\ \bibinfo {author} {\bibfnamefont {C.}~\bibnamefont {von Borczyskowski}},\ }\bibfield  {title} {\bibinfo {title} {Scanning confocal optical microscopy and magnetic resonance on single defect centers},\ }\href {https://doi.org/10.1126/science.276.5321.2012} {\bibfield  {journal} {\bibinfo  {journal} {Science}\ }\textbf {\bibinfo {volume} {276}},\ \bibinfo {pages} {2012} (\bibinfo {year} {1997})}\BibitemShut {NoStop}%
\bibitem [{\citenamefont {Dr\"abenstedt}\ \emph {et~al.}(1999)\citenamefont {Dr\"abenstedt}, \citenamefont {Fleury}, \citenamefont {Tietz}, \citenamefont {Jelezko}, \citenamefont {Kilin}, \citenamefont {Nizovtzev},\ and\ \citenamefont {Wrachtrup}}]{PhysRevB.60.11503}%
  \BibitemOpen
  \bibfield  {author} {\bibinfo {author} {\bibfnamefont {A.}~\bibnamefont {Dr\"abenstedt}}, \bibinfo {author} {\bibfnamefont {L.}~\bibnamefont {Fleury}}, \bibinfo {author} {\bibfnamefont {C.}~\bibnamefont {Tietz}}, \bibinfo {author} {\bibfnamefont {F.}~\bibnamefont {Jelezko}}, \bibinfo {author} {\bibfnamefont {S.}~\bibnamefont {Kilin}}, \bibinfo {author} {\bibfnamefont {A.}~\bibnamefont {Nizovtzev}},\ and\ \bibinfo {author} {\bibfnamefont {J.}~\bibnamefont {Wrachtrup}},\ }\bibfield  {title} {\bibinfo {title} {Low-temperature microscopy and spectroscopy on single defect centers in diamond},\ }\href {https://doi.org/10.1103/PhysRevB.60.11503} {\bibfield  {journal} {\bibinfo  {journal} {Phys. Rev. B}\ }\textbf {\bibinfo {volume} {60}},\ \bibinfo {pages} {11503} (\bibinfo {year} {1999})}\BibitemShut {NoStop}%
\bibitem [{\citenamefont {Jelezko}\ \emph {et~al.}(2001)\citenamefont {Jelezko}, \citenamefont {Tietz}, \citenamefont {Gruber}, \citenamefont {Popa}, \citenamefont {Nizovtsev}, \citenamefont {Kilin},\ and\ \citenamefont {Wrachtrup}}]{Jele01}%
  \BibitemOpen
  \bibfield  {author} {\bibinfo {author} {\bibfnamefont {F.}~\bibnamefont {Jelezko}}, \bibinfo {author} {\bibfnamefont {C.}~\bibnamefont {Tietz}}, \bibinfo {author} {\bibfnamefont {A.}~\bibnamefont {Gruber}}, \bibinfo {author} {\bibfnamefont {I.}~\bibnamefont {Popa}}, \bibinfo {author} {\bibfnamefont {A.}~\bibnamefont {Nizovtsev}}, \bibinfo {author} {\bibfnamefont {S.}~\bibnamefont {Kilin}},\ and\ \bibinfo {author} {\bibfnamefont {J.}~\bibnamefont {Wrachtrup}},\ }\bibfield  {title} {\bibinfo {title} {Spectroscopy of single $\text{N-V}$ centers in diamond},\ }\href {https://doi.org/https://doi.org/10.1002/1438-5171(200112)2:4<255::AID-SIMO255>3.0.CO;2-D} {\bibfield  {journal} {\bibinfo  {journal} {Single Molecules}\ }\textbf {\bibinfo {volume} {2}},\ \bibinfo {pages} {255} (\bibinfo {year} {2001})}\BibitemShut {NoStop}%
\bibitem [{\citenamefont {Chernobrod}\ and\ \citenamefont {Berman}(2004)}]{Cher04}%
  \BibitemOpen
  \bibfield  {author} {\bibinfo {author} {\bibfnamefont {B.~M.}\ \bibnamefont {Chernobrod}}\ and\ \bibinfo {author} {\bibfnamefont {G.~P.}\ \bibnamefont {Berman}},\ }\bibfield  {title} {\bibinfo {title} {Spin microscope based on optically detected magnetic resonance},\ }\href {https://doi.org/10.1063/1.1829373} {\bibfield  {journal} {\bibinfo  {journal} {Journal of Applied Physics}\ }\textbf {\bibinfo {volume} {97}},\ \bibinfo {pages} {014903} (\bibinfo {year} {2004})}\BibitemShut {NoStop}%
\bibitem [{\citenamefont {Balasubramanian}\ \emph {et~al.}(2008)\citenamefont {Balasubramanian}, \citenamefont {Chan}, \citenamefont {Kolesov}, \citenamefont {Al-Hmoud}, \citenamefont {Tisler}, \citenamefont {Shin}, \citenamefont {Kim}, \citenamefont {Wojcik}, \citenamefont {Hemmer}, \citenamefont {Krueger}, \citenamefont {Hanke}, \citenamefont {Leitenstorfer}, \citenamefont {Bratschitsch}, \citenamefont {Jelezko},\ and\ \citenamefont {Wrachtrup}}]{Bala08}%
  \BibitemOpen
  \bibfield  {author} {\bibinfo {author} {\bibfnamefont {G.}~\bibnamefont {Balasubramanian}}, \bibinfo {author} {\bibfnamefont {I.~Y.}\ \bibnamefont {Chan}}, \bibinfo {author} {\bibfnamefont {R.}~\bibnamefont {Kolesov}}, \bibinfo {author} {\bibfnamefont {M.}~\bibnamefont {Al-Hmoud}}, \bibinfo {author} {\bibfnamefont {J.}~\bibnamefont {Tisler}}, \bibinfo {author} {\bibfnamefont {C.}~\bibnamefont {Shin}}, \bibinfo {author} {\bibfnamefont {C.}~\bibnamefont {Kim}}, \bibinfo {author} {\bibfnamefont {A.}~\bibnamefont {Wojcik}}, \bibinfo {author} {\bibfnamefont {P.}~\bibnamefont {Hemmer}}, \bibinfo {author} {\bibfnamefont {A.}~\bibnamefont {Krueger}}, \bibinfo {author} {\bibfnamefont {T.}~\bibnamefont {Hanke}}, \bibinfo {author} {\bibfnamefont {A.}~\bibnamefont {Leitenstorfer}}, \bibinfo {author} {\bibfnamefont {R.}~\bibnamefont {Bratschitsch}}, \bibinfo {author} {\bibfnamefont {F.}~\bibnamefont {Jelezko}},\ and\ \bibinfo {author} {\bibfnamefont {J.}~\bibnamefont {Wrachtrup}},\ }\bibfield  {title} {\bibinfo {title}
  {Nanoscale imaging magnetometry with diamond spins under ambient conditions},\ }\href {https://doi.org/10.1038/nature07278.} {\bibfield  {journal} {\bibinfo  {journal} {Nature}\ }\textbf {\bibinfo {volume} {455}},\ \bibinfo {pages} {648} (\bibinfo {year} {2008})}\BibitemShut {NoStop}%
\bibitem [{\citenamefont {Maze}\ \emph {et~al.}(2008)\citenamefont {Maze}, \citenamefont {Stanwix}, \citenamefont {Hodges}, \citenamefont {Hong}, \citenamefont {Taylor}, \citenamefont {Cappellaro}, \citenamefont {Jiang}, \citenamefont {Dutt}, \citenamefont {Togan}, \citenamefont {Zibrov}, \citenamefont {Yacoby}, \citenamefont {Walsworth},\ and\ \citenamefont {Lukin}}]{Maze08}%
  \BibitemOpen
  \bibfield  {author} {\bibinfo {author} {\bibfnamefont {J.~R.}\ \bibnamefont {Maze}}, \bibinfo {author} {\bibfnamefont {P.~L.}\ \bibnamefont {Stanwix}}, \bibinfo {author} {\bibfnamefont {J.~S.}\ \bibnamefont {Hodges}}, \bibinfo {author} {\bibfnamefont {S.}~\bibnamefont {Hong}}, \bibinfo {author} {\bibfnamefont {J.~M.}\ \bibnamefont {Taylor}}, \bibinfo {author} {\bibfnamefont {P.}~\bibnamefont {Cappellaro}}, \bibinfo {author} {\bibfnamefont {L.}~\bibnamefont {Jiang}}, \bibinfo {author} {\bibfnamefont {M.~V.~G.}\ \bibnamefont {Dutt}}, \bibinfo {author} {\bibfnamefont {E.}~\bibnamefont {Togan}}, \bibinfo {author} {\bibfnamefont {A.~S.}\ \bibnamefont {Zibrov}}, \bibinfo {author} {\bibfnamefont {A.}~\bibnamefont {Yacoby}}, \bibinfo {author} {\bibfnamefont {R.~L.}\ \bibnamefont {Walsworth}},\ and\ \bibinfo {author} {\bibfnamefont {M.~D.}\ \bibnamefont {Lukin}},\ }\bibfield  {title} {\bibinfo {title} {Nanoscale magnetic sensing with an individual electronic spin in diamond},\ }\href
  {https://doi.org/https://doi.org/10.1038/nature07279} {\bibfield  {journal} {\bibinfo  {journal} {Nature}\ }\textbf {\bibinfo {volume} {455}},\ \bibinfo {pages} {644} (\bibinfo {year} {2008})}\BibitemShut {NoStop}%
\bibitem [{\citenamefont {Taylor}\ \emph {et~al.}(2008)\citenamefont {Taylor}, \citenamefont {Cappellaro}, \citenamefont {Childress}, \citenamefont {Jiang}, \citenamefont {Budker}, \citenamefont {Hemmer}, \citenamefont {Yacoby}, \citenamefont {Walsworth},\ and\ \citenamefont {Lukin}}]{Tayl08}%
  \BibitemOpen
  \bibfield  {author} {\bibinfo {author} {\bibfnamefont {J.~M.}\ \bibnamefont {Taylor}}, \bibinfo {author} {\bibfnamefont {P.}~\bibnamefont {Cappellaro}}, \bibinfo {author} {\bibfnamefont {L.}~\bibnamefont {Childress}}, \bibinfo {author} {\bibfnamefont {L.}~\bibnamefont {Jiang}}, \bibinfo {author} {\bibfnamefont {D.}~\bibnamefont {Budker}}, \bibinfo {author} {\bibfnamefont {P.~R.}\ \bibnamefont {Hemmer}}, \bibinfo {author} {\bibfnamefont {A.}~\bibnamefont {Yacoby}}, \bibinfo {author} {\bibfnamefont {R.}~\bibnamefont {Walsworth}},\ and\ \bibinfo {author} {\bibfnamefont {M.~D.}\ \bibnamefont {Lukin}},\ }\bibfield  {title} {\bibinfo {title} {High-sensitivity diamond magnetometer with nanoscale resolution},\ }\href {https://doi.org/https://doi.org/10.1038/nphys1075} {\bibfield  {journal} {\bibinfo  {journal} {High-sensitivity diamond magnetometer with nanoscale resolution}\ }\textbf {\bibinfo {volume} {4}},\ \bibinfo {pages} {810} (\bibinfo {year} {2008})}\BibitemShut {NoStop}%
\bibitem [{\citenamefont {Mamin}\ \emph {et~al.}(2013)\citenamefont {Mamin}, \citenamefont {Kim}, \citenamefont {Sherwood}, \citenamefont {Rettner}, \citenamefont {Ohno}, \citenamefont {Awschalom},\ and\ \citenamefont {Rugar}}]{Mami13}%
  \BibitemOpen
  \bibfield  {author} {\bibinfo {author} {\bibfnamefont {H.~J.}\ \bibnamefont {Mamin}}, \bibinfo {author} {\bibfnamefont {M.}~\bibnamefont {Kim}}, \bibinfo {author} {\bibfnamefont {M.~H.}\ \bibnamefont {Sherwood}}, \bibinfo {author} {\bibfnamefont {C.~T.}\ \bibnamefont {Rettner}}, \bibinfo {author} {\bibfnamefont {K.}~\bibnamefont {Ohno}}, \bibinfo {author} {\bibfnamefont {D.~D.}\ \bibnamefont {Awschalom}},\ and\ \bibinfo {author} {\bibfnamefont {D.}~\bibnamefont {Rugar}},\ }\bibfield  {title} {\bibinfo {title} {Nanoscale nuclear magnetic resonance with a nitrogen-vacancy spin sensor},\ }\href {https://doi.org/10.1126/science.1231540} {\bibfield  {journal} {\bibinfo  {journal} {Science}\ }\textbf {\bibinfo {volume} {339}},\ \bibinfo {pages} {557} (\bibinfo {year} {2013})}\BibitemShut {NoStop}%
\bibitem [{\citenamefont {Staudacher}\ \emph {et~al.}(2013)\citenamefont {Staudacher}, \citenamefont {Shi}, \citenamefont {Pezzagna}, \citenamefont {Meijer}, \citenamefont {Du}, \citenamefont {Meriles}, \citenamefont {Reinhard},\ and\ \citenamefont {Wrachtrup}}]{Stau13}%
  \BibitemOpen
  \bibfield  {author} {\bibinfo {author} {\bibfnamefont {T.}~\bibnamefont {Staudacher}}, \bibinfo {author} {\bibfnamefont {F.}~\bibnamefont {Shi}}, \bibinfo {author} {\bibfnamefont {S.}~\bibnamefont {Pezzagna}}, \bibinfo {author} {\bibfnamefont {J.}~\bibnamefont {Meijer}}, \bibinfo {author} {\bibfnamefont {J.}~\bibnamefont {Du}}, \bibinfo {author} {\bibfnamefont {C.~A.}\ \bibnamefont {Meriles}}, \bibinfo {author} {\bibfnamefont {F.}~\bibnamefont {Reinhard}},\ and\ \bibinfo {author} {\bibfnamefont {J.}~\bibnamefont {Wrachtrup}},\ }\bibfield  {title} {\bibinfo {title} {Nuclear magnetic resonance spectroscopy on a (5-nanometer)$^3$ sample volume},\ }\href {https://doi.org/10.1126/science.1231675} {\bibfield  {journal} {\bibinfo  {journal} {Science}\ }\textbf {\bibinfo {volume} {339}},\ \bibinfo {pages} {561} (\bibinfo {year} {2013})}\BibitemShut {NoStop}%
\bibitem [{\citenamefont {Maertz}\ \emph {et~al.}(2010)\citenamefont {Maertz}, \citenamefont {Wijnheijmer}, \citenamefont {Fuchs}, \citenamefont {Nowakowski},\ and\ \citenamefont {Awschalom}}]{Maer10}%
  \BibitemOpen
  \bibfield  {author} {\bibinfo {author} {\bibfnamefont {B.~J.}\ \bibnamefont {Maertz}}, \bibinfo {author} {\bibfnamefont {A.~P.}\ \bibnamefont {Wijnheijmer}}, \bibinfo {author} {\bibfnamefont {G.~D.}\ \bibnamefont {Fuchs}}, \bibinfo {author} {\bibfnamefont {M.~E.}\ \bibnamefont {Nowakowski}},\ and\ \bibinfo {author} {\bibfnamefont {D.~D.}\ \bibnamefont {Awschalom}},\ }\bibfield  {title} {\bibinfo {title} {Vector magnetic field microscopy using nitrogen vacancy centers in diamond},\ }\href@noop {} {\bibfield  {journal} {\bibinfo  {journal} {Applied Physics Letters}\ }\textbf {\bibinfo {volume} {96}} (\bibinfo {year} {2010})}\BibitemShut {NoStop}%
\bibitem [{\citenamefont {Steinert}\ \emph {et~al.}(2010)\citenamefont {Steinert}, \citenamefont {Dolde}, \citenamefont {Neumann}, \citenamefont {Aird}, \citenamefont {Naydenov}, \citenamefont {Balasubramanian}, \citenamefont {Jelezko},\ and\ \citenamefont {Wrachtrup}}]{Stei10}%
  \BibitemOpen
  \bibfield  {author} {\bibinfo {author} {\bibfnamefont {S.}~\bibnamefont {Steinert}}, \bibinfo {author} {\bibfnamefont {F.}~\bibnamefont {Dolde}}, \bibinfo {author} {\bibfnamefont {P.}~\bibnamefont {Neumann}}, \bibinfo {author} {\bibfnamefont {A.}~\bibnamefont {Aird}}, \bibinfo {author} {\bibfnamefont {B.}~\bibnamefont {Naydenov}}, \bibinfo {author} {\bibfnamefont {G.}~\bibnamefont {Balasubramanian}}, \bibinfo {author} {\bibfnamefont {F.}~\bibnamefont {Jelezko}},\ and\ \bibinfo {author} {\bibfnamefont {J.}~\bibnamefont {Wrachtrup}},\ }\bibfield  {title} {\bibinfo {title} {High sensitivity magnetic imaging using an array of spins in diamond},\ }\href@noop {} {\bibfield  {journal} {\bibinfo  {journal} {Review of scientific instruments}\ }\textbf {\bibinfo {volume} {81}} (\bibinfo {year} {2010})}\BibitemShut {NoStop}%
\bibitem [{\citenamefont {Michl}\ \emph {et~al.}(2014)\citenamefont {Michl}, \citenamefont {Teraji}, \citenamefont {Zaiser}, \citenamefont {Jakobi}, \citenamefont {Waldherr}, \citenamefont {Dolde}, \citenamefont {Neumann}, \citenamefont {Doherty}, \citenamefont {Manson}, \citenamefont {Isoya},\ and\ \citenamefont {Wrachtrup}}]{mich14}%
  \BibitemOpen
  \bibfield  {author} {\bibinfo {author} {\bibfnamefont {J.}~\bibnamefont {Michl}}, \bibinfo {author} {\bibfnamefont {T.}~\bibnamefont {Teraji}}, \bibinfo {author} {\bibfnamefont {S.}~\bibnamefont {Zaiser}}, \bibinfo {author} {\bibfnamefont {I.}~\bibnamefont {Jakobi}}, \bibinfo {author} {\bibfnamefont {G.}~\bibnamefont {Waldherr}}, \bibinfo {author} {\bibfnamefont {F.}~\bibnamefont {Dolde}}, \bibinfo {author} {\bibfnamefont {P.}~\bibnamefont {Neumann}}, \bibinfo {author} {\bibfnamefont {M.~W.}\ \bibnamefont {Doherty}}, \bibinfo {author} {\bibfnamefont {N.~B.}\ \bibnamefont {Manson}}, \bibinfo {author} {\bibfnamefont {J.}~\bibnamefont {Isoya}},\ and\ \bibinfo {author} {\bibfnamefont {J.}~\bibnamefont {Wrachtrup}},\ }\bibfield  {title} {\bibinfo {title} {Perfect alignment and preferential orientation of nitrogen-vacancy centers during chemical vapor deposition diamond growth on (111) surfaces},\ }\href@noop {} {\bibfield  {journal} {\bibinfo  {journal} {Applied Physics Letters}\ }\textbf {\bibinfo {volume}
  {104}} (\bibinfo {year} {2014})}\BibitemShut {NoStop}%
\bibitem [{\citenamefont {Lesik}\ \emph {et~al.}(2014)\citenamefont {Lesik}, \citenamefont {Tetienne}, \citenamefont {Tallaire}, \citenamefont {Achard}, \citenamefont {Mille}, \citenamefont {Gicquel}, \citenamefont {Roch},\ and\ \citenamefont {Jacques}}]{lesi14}%
  \BibitemOpen
  \bibfield  {author} {\bibinfo {author} {\bibfnamefont {M.}~\bibnamefont {Lesik}}, \bibinfo {author} {\bibfnamefont {J.-P.}\ \bibnamefont {Tetienne}}, \bibinfo {author} {\bibfnamefont {A.}~\bibnamefont {Tallaire}}, \bibinfo {author} {\bibfnamefont {J.}~\bibnamefont {Achard}}, \bibinfo {author} {\bibfnamefont {V.}~\bibnamefont {Mille}}, \bibinfo {author} {\bibfnamefont {A.}~\bibnamefont {Gicquel}}, \bibinfo {author} {\bibfnamefont {J.-F.}\ \bibnamefont {Roch}},\ and\ \bibinfo {author} {\bibfnamefont {V.}~\bibnamefont {Jacques}},\ }\bibfield  {title} {\bibinfo {title} {Perfect preferential orientation of nitrogen-vacancy defects in a synthetic diamond sample},\ }\href@noop {} {\bibfield  {journal} {\bibinfo  {journal} {Applied Physics Letters}\ }\textbf {\bibinfo {volume} {104}} (\bibinfo {year} {2014})}\BibitemShut {NoStop}%
\bibitem [{\citenamefont {Tetienne}\ \emph {et~al.}(2012)\citenamefont {Tetienne}, \citenamefont {Rondin}, \citenamefont {Spinicelli}, \citenamefont {Chipaux}, \citenamefont {Debuisschert}, \citenamefont {Roch},\ and\ \citenamefont {Jacques}}]{Teti12}%
  \BibitemOpen
  \bibfield  {author} {\bibinfo {author} {\bibfnamefont {J.~P.}\ \bibnamefont {Tetienne}}, \bibinfo {author} {\bibfnamefont {L.}~\bibnamefont {Rondin}}, \bibinfo {author} {\bibfnamefont {P.}~\bibnamefont {Spinicelli}}, \bibinfo {author} {\bibfnamefont {M.}~\bibnamefont {Chipaux}}, \bibinfo {author} {\bibfnamefont {T.}~\bibnamefont {Debuisschert}}, \bibinfo {author} {\bibfnamefont {J.-F.}\ \bibnamefont {Roch}},\ and\ \bibinfo {author} {\bibfnamefont {V.}~\bibnamefont {Jacques}},\ }\bibfield  {title} {\bibinfo {title} {Magnetic-field-dependent photodynamics of single $\text{NV}$ defects in diamond: an application to qualitative all-optical magnetic imaging},\ }\href@noop {} {\bibfield  {journal} {\bibinfo  {journal} {New Journal of Physics}\ }\textbf {\bibinfo {volume} {14}},\ \bibinfo {pages} {103033} (\bibinfo {year} {2012})}\BibitemShut {NoStop}%
\bibitem [{\citenamefont {Simin}\ \emph {et~al.}(2015)\citenamefont {Simin}, \citenamefont {Fuchs}, \citenamefont {Kraus}, \citenamefont {Sperlich}, \citenamefont {Baranov}, \citenamefont {Astakhov},\ and\ \citenamefont {Dyakonov}}]{simin2015high}%
  \BibitemOpen
  \bibfield  {author} {\bibinfo {author} {\bibfnamefont {D.}~\bibnamefont {Simin}}, \bibinfo {author} {\bibfnamefont {F.}~\bibnamefont {Fuchs}}, \bibinfo {author} {\bibfnamefont {H.}~\bibnamefont {Kraus}}, \bibinfo {author} {\bibfnamefont {A.}~\bibnamefont {Sperlich}}, \bibinfo {author} {\bibfnamefont {P.~G.}\ \bibnamefont {Baranov}}, \bibinfo {author} {\bibfnamefont {G.~V.}\ \bibnamefont {Astakhov}},\ and\ \bibinfo {author} {\bibfnamefont {V.}~\bibnamefont {Dyakonov}},\ }\bibfield  {title} {\bibinfo {title} {High-precision angle-resolved magnetometry with uniaxial quantum centers in silicon carbide},\ }\href@noop {} {\bibfield  {journal} {\bibinfo  {journal} {Physical Review Applied}\ }\textbf {\bibinfo {volume} {4}},\ \bibinfo {pages} {014009} (\bibinfo {year} {2015})}\BibitemShut {NoStop}%
\bibitem [{\citenamefont {Kraus}\ \emph {et~al.}(2014)\citenamefont {Kraus}, \citenamefont {Soltamov}, \citenamefont {Fuchs}, \citenamefont {Simin}, \citenamefont {Sperlich}, \citenamefont {Baranov}, \citenamefont {Astakhov},\ and\ \citenamefont {Dyakonov}}]{kraus2014magnetic}%
  \BibitemOpen
  \bibfield  {author} {\bibinfo {author} {\bibfnamefont {H.}~\bibnamefont {Kraus}}, \bibinfo {author} {\bibfnamefont {V.~A.}\ \bibnamefont {Soltamov}}, \bibinfo {author} {\bibfnamefont {F.}~\bibnamefont {Fuchs}}, \bibinfo {author} {\bibfnamefont {D.}~\bibnamefont {Simin}}, \bibinfo {author} {\bibfnamefont {A.}~\bibnamefont {Sperlich}}, \bibinfo {author} {\bibfnamefont {P.~G.}\ \bibnamefont {Baranov}}, \bibinfo {author} {\bibfnamefont {G.~V.}\ \bibnamefont {Astakhov}},\ and\ \bibinfo {author} {\bibfnamefont {V.}~\bibnamefont {Dyakonov}},\ }\bibfield  {title} {\bibinfo {title} {Magnetic field and temperature sensing with atomic-scale spin defects in silicon carbide},\ }\href@noop {} {\bibfield  {journal} {\bibinfo  {journal} {Scientific reports}\ }\textbf {\bibinfo {volume} {4}},\ \bibinfo {pages} {5303} (\bibinfo {year} {2014})}\BibitemShut {NoStop}%
\bibitem [{\citenamefont {Baranov}\ \emph {et~al.}(2007)\citenamefont {Baranov}, \citenamefont {Bundakova}, \citenamefont {Borovykh}, \citenamefont {Orlinski\u\i}, \citenamefont {Zondervan},\ and\ \citenamefont {Schmidt}}]{baranov2007spin}%
  \BibitemOpen
  \bibfield  {author} {\bibinfo {author} {\bibfnamefont {P.~G.}\ \bibnamefont {Baranov}}, \bibinfo {author} {\bibfnamefont {A.~P.}\ \bibnamefont {Bundakova}}, \bibinfo {author} {\bibfnamefont {I.}~\bibnamefont {Borovykh}}, \bibinfo {author} {\bibfnamefont {S.}~\bibnamefont {Orlinski\u\i}}, \bibinfo {author} {\bibfnamefont {R.}~\bibnamefont {Zondervan}},\ and\ \bibinfo {author} {\bibfnamefont {J.}~\bibnamefont {Schmidt}},\ }\bibfield  {title} {\bibinfo {title} {Spin polarization induced by optical and microwave resonance radiation in a $\text{Si}$ vacancy in $\text{SiC}$: A promising subject for the spectroscopy of single defects},\ }\href@noop {} {\bibfield  {journal} {\bibinfo  {journal} {JETP Letters}\ }\textbf {\bibinfo {volume} {86}},\ \bibinfo {pages} {202} (\bibinfo {year} {2007})}\BibitemShut {NoStop}%
\bibitem [{\citenamefont {Baranov}\ \emph {et~al.}(2011)\citenamefont {Baranov}, \citenamefont {Bundakova}, \citenamefont {Soltamova}, \citenamefont {Orlinskii}, \citenamefont {Borovykh}, \citenamefont {Zondervan}, \citenamefont {Verberk},\ and\ \citenamefont {Schmidt}}]{baranov2011silicon}%
  \BibitemOpen
  \bibfield  {author} {\bibinfo {author} {\bibfnamefont {P.~G.}\ \bibnamefont {Baranov}}, \bibinfo {author} {\bibfnamefont {A.~P.}\ \bibnamefont {Bundakova}}, \bibinfo {author} {\bibfnamefont {A.~A.}\ \bibnamefont {Soltamova}}, \bibinfo {author} {\bibfnamefont {S.~B.}\ \bibnamefont {Orlinskii}}, \bibinfo {author} {\bibfnamefont {I.~V.}\ \bibnamefont {Borovykh}}, \bibinfo {author} {\bibfnamefont {R.}~\bibnamefont {Zondervan}}, \bibinfo {author} {\bibfnamefont {R.}~\bibnamefont {Verberk}},\ and\ \bibinfo {author} {\bibfnamefont {J.}~\bibnamefont {Schmidt}},\ }\bibfield  {title} {\bibinfo {title} {Silicon vacancy in $\text{SiC}$ as a promising quantum system for single-defect and single-photon spectroscopy},\ }\href@noop {} {\bibfield  {journal} {\bibinfo  {journal} {Physical Review B}\ }\textbf {\bibinfo {volume} {83}},\ \bibinfo {pages} {125203} (\bibinfo {year} {2011})}\BibitemShut {NoStop}%
\bibitem [{\citenamefont {Soltamov}\ \emph {et~al.}(2015)\citenamefont {Soltamov}, \citenamefont {Yavkin}, \citenamefont {Tolmachev}, \citenamefont {Babunts}, \citenamefont {Badalyan}, \citenamefont {Davydov}, \citenamefont {Mokhov}, \citenamefont {Proskuryakov}, \citenamefont {Orlinskii},\ and\ \citenamefont {Baranov}}]{soltamov2015optically}%
  \BibitemOpen
  \bibfield  {author} {\bibinfo {author} {\bibfnamefont {V.~A.}\ \bibnamefont {Soltamov}}, \bibinfo {author} {\bibfnamefont {B.~V.}\ \bibnamefont {Yavkin}}, \bibinfo {author} {\bibfnamefont {D.~O.}\ \bibnamefont {Tolmachev}}, \bibinfo {author} {\bibfnamefont {R.~A.}\ \bibnamefont {Babunts}}, \bibinfo {author} {\bibfnamefont {A.~G.}\ \bibnamefont {Badalyan}}, \bibinfo {author} {\bibfnamefont {V.~Y.}\ \bibnamefont {Davydov}}, \bibinfo {author} {\bibfnamefont {E.~N.}\ \bibnamefont {Mokhov}}, \bibinfo {author} {\bibfnamefont {I.~I.}\ \bibnamefont {Proskuryakov}}, \bibinfo {author} {\bibfnamefont {S.~B.}\ \bibnamefont {Orlinskii}},\ and\ \bibinfo {author} {\bibfnamefont {P.~G.}\ \bibnamefont {Baranov}},\ }\bibfield  {title} {\bibinfo {title} {Optically addressable silicon vacancy-related spin centers in rhombic silicon carbide with high breakdown characteristics and endor evidence of their structure},\ }\href@noop {} {\bibfield  {journal} {\bibinfo  {journal} {Physical review letters}\ }\textbf {\bibinfo {volume}
  {115}},\ \bibinfo {pages} {247602} (\bibinfo {year} {2015})}\BibitemShut {NoStop}%
\bibitem [{\citenamefont {Lee}\ \emph {et~al.}(2015)\citenamefont {Lee}, \citenamefont {Niethammer},\ and\ \citenamefont {Wrachtrup}}]{lee2015vector}%
  \BibitemOpen
  \bibfield  {author} {\bibinfo {author} {\bibfnamefont {S.-Y.}\ \bibnamefont {Lee}}, \bibinfo {author} {\bibfnamefont {M.}~\bibnamefont {Niethammer}},\ and\ \bibinfo {author} {\bibfnamefont {J.}~\bibnamefont {Wrachtrup}},\ }\bibfield  {title} {\bibinfo {title} {Vector magnetometry based on $\textit{S}=3/2$ electronic spins},\ }\href@noop {} {\bibfield  {journal} {\bibinfo  {journal} {Physical Review B}\ }\textbf {\bibinfo {volume} {92}},\ \bibinfo {pages} {115201} (\bibinfo {year} {2015})}\BibitemShut {NoStop}%
\bibitem [{\citenamefont {Niethammer}\ \emph {et~al.}(2016)\citenamefont {Niethammer}, \citenamefont {Widmann}, \citenamefont {Lee}, \citenamefont {Stenberg}, \citenamefont {Kordina}, \citenamefont {Ohshima}, \citenamefont {Son}, \citenamefont {Janz{\'e}n},\ and\ \citenamefont {Wrachtrup}}]{niethammer2016vector}%
  \BibitemOpen
  \bibfield  {author} {\bibinfo {author} {\bibfnamefont {M.}~\bibnamefont {Niethammer}}, \bibinfo {author} {\bibfnamefont {M.}~\bibnamefont {Widmann}}, \bibinfo {author} {\bibfnamefont {S.-Y.}\ \bibnamefont {Lee}}, \bibinfo {author} {\bibfnamefont {P.}~\bibnamefont {Stenberg}}, \bibinfo {author} {\bibfnamefont {O.}~\bibnamefont {Kordina}}, \bibinfo {author} {\bibfnamefont {T.}~\bibnamefont {Ohshima}}, \bibinfo {author} {\bibfnamefont {N.~T.}\ \bibnamefont {Son}}, \bibinfo {author} {\bibfnamefont {E.}~\bibnamefont {Janz{\'e}n}},\ and\ \bibinfo {author} {\bibfnamefont {J.}~\bibnamefont {Wrachtrup}},\ }\bibfield  {title} {\bibinfo {title} {Vector magnetometry using silicon vacancies in $\text{4H-SiC}$ under ambient conditions},\ }\href@noop {} {\bibfield  {journal} {\bibinfo  {journal} {Physical Review Applied}\ }\textbf {\bibinfo {volume} {6}},\ \bibinfo {pages} {034001} (\bibinfo {year} {2016})}\BibitemShut {NoStop}%
\bibitem [{\citenamefont {Anisimov}\ \emph {et~al.}(2016{\natexlab{a}})\citenamefont {Anisimov}, \citenamefont {Babunts}, \citenamefont {Kidalov}, \citenamefont {Mokhov}, \citenamefont {Soltamov},\ and\ \citenamefont {Baranov}}]{anisimov2016spin}%
  \BibitemOpen
  \bibfield  {author} {\bibinfo {author} {\bibfnamefont {A.~N.}\ \bibnamefont {Anisimov}}, \bibinfo {author} {\bibfnamefont {R.~A.}\ \bibnamefont {Babunts}}, \bibinfo {author} {\bibfnamefont {S.~V.}\ \bibnamefont {Kidalov}}, \bibinfo {author} {\bibfnamefont {E.~N.}\ \bibnamefont {Mokhov}}, \bibinfo {author} {\bibfnamefont {V.~A.}\ \bibnamefont {Soltamov}},\ and\ \bibinfo {author} {\bibfnamefont {P.~G.}\ \bibnamefont {Baranov}},\ }\bibfield  {title} {\bibinfo {title} {Spin centers in $\text{SiC}$ for all-optical nanoscale quantum sensing under ambient conditions},\ }\href@noop {} {\bibfield  {journal} {\bibinfo  {journal} {JETP letters}\ }\textbf {\bibinfo {volume} {104}},\ \bibinfo {pages} {82} (\bibinfo {year} {2016}{\natexlab{a}})}\BibitemShut {NoStop}%
\bibitem [{\citenamefont {Simin}\ \emph {et~al.}(2016)\citenamefont {Simin}, \citenamefont {Soltamov}, \citenamefont {Poshakinskiy}, \citenamefont {Anisimov}, \citenamefont {Babunts}, \citenamefont {Tolmachev}, \citenamefont {Mokhov}, \citenamefont {Trupke}, \citenamefont {Tarasenko}, \citenamefont {Sperlich}, \citenamefont {Baranov}, \citenamefont {Dyakonov},\ and\ \citenamefont {Astakhov}}]{simin2016all}%
  \BibitemOpen
  \bibfield  {author} {\bibinfo {author} {\bibfnamefont {D.}~\bibnamefont {Simin}}, \bibinfo {author} {\bibfnamefont {V.~A.}\ \bibnamefont {Soltamov}}, \bibinfo {author} {\bibfnamefont {A.~V.}\ \bibnamefont {Poshakinskiy}}, \bibinfo {author} {\bibfnamefont {A.~N.}\ \bibnamefont {Anisimov}}, \bibinfo {author} {\bibfnamefont {R.~A.}\ \bibnamefont {Babunts}}, \bibinfo {author} {\bibfnamefont {D.~O.}\ \bibnamefont {Tolmachev}}, \bibinfo {author} {\bibfnamefont {E.~N.}\ \bibnamefont {Mokhov}}, \bibinfo {author} {\bibfnamefont {M.}~\bibnamefont {Trupke}}, \bibinfo {author} {\bibfnamefont {S.~A.}\ \bibnamefont {Tarasenko}}, \bibinfo {author} {\bibfnamefont {A.}~\bibnamefont {Sperlich}}, \bibinfo {author} {\bibfnamefont {P.~G.}\ \bibnamefont {Baranov}}, \bibinfo {author} {\bibfnamefont {V.}~\bibnamefont {Dyakonov}},\ and\ \bibinfo {author} {\bibfnamefont {G.~V.}\ \bibnamefont {Astakhov}},\ }\bibfield  {title} {\bibinfo {title} {All-optical dc nanotesla magnetometry using silicon vacancy fine structure in isotopically
  purified silicon carbide},\ }\href@noop {} {\bibfield  {journal} {\bibinfo  {journal} {Physical Review X}\ }\textbf {\bibinfo {volume} {6}},\ \bibinfo {pages} {031014} (\bibinfo {year} {2016})}\BibitemShut {NoStop}%
\bibitem [{\citenamefont {Anisimov}\ \emph {et~al.}(2016{\natexlab{b}})\citenamefont {Anisimov}, \citenamefont {Simin}, \citenamefont {Soltamov}, \citenamefont {Lebedev}, \citenamefont {Baranov}, \citenamefont {Astakhov},\ and\ \citenamefont {Dyakonov}}]{anisimov2016optical}%
  \BibitemOpen
  \bibfield  {author} {\bibinfo {author} {\bibfnamefont {A.~N.}\ \bibnamefont {Anisimov}}, \bibinfo {author} {\bibfnamefont {D.}~\bibnamefont {Simin}}, \bibinfo {author} {\bibfnamefont {V.~A.}\ \bibnamefont {Soltamov}}, \bibinfo {author} {\bibfnamefont {S.~P.}\ \bibnamefont {Lebedev}}, \bibinfo {author} {\bibfnamefont {P.~G.}\ \bibnamefont {Baranov}}, \bibinfo {author} {\bibfnamefont {G.~V.}\ \bibnamefont {Astakhov}},\ and\ \bibinfo {author} {\bibfnamefont {V.}~\bibnamefont {Dyakonov}},\ }\bibfield  {title} {\bibinfo {title} {Optical thermometry based on level anticrossing in silicon carbide},\ }\href@noop {} {\bibfield  {journal} {\bibinfo  {journal} {Scientific reports}\ }\textbf {\bibinfo {volume} {6}},\ \bibinfo {pages} {33301} (\bibinfo {year} {2016}{\natexlab{b}})}\BibitemShut {NoStop}%
\bibitem [{\citenamefont {Baranov}\ \emph {et~al.}(2017{\natexlab{a}})\citenamefont {Baranov}, \citenamefont {Von~Bardeleben}, \citenamefont {Jelezko},\ and\ \citenamefont {Wrachtrup}}]{baranov2017magnetic}%
  \BibitemOpen
  \bibfield  {author} {\bibinfo {author} {\bibfnamefont {P.~G.}\ \bibnamefont {Baranov}}, \bibinfo {author} {\bibfnamefont {H.~J.}\ \bibnamefont {Von~Bardeleben}}, \bibinfo {author} {\bibfnamefont {F.}~\bibnamefont {Jelezko}},\ and\ \bibinfo {author} {\bibfnamefont {J.}~\bibnamefont {Wrachtrup}},\ }\href@noop {} {\emph {\bibinfo {title} {Magnetic Resonance of Semiconductors and Their Nanostructures}}}\ (\bibinfo  {publisher} {Springer},\ \bibinfo {year} {2017})\BibitemShut {NoStop}%
\bibitem [{\citenamefont {Baranov}\ \emph {et~al.}(2017{\natexlab{b}})\citenamefont {Baranov}, \citenamefont {Von~Bardeleben}, \citenamefont {Jelezko},\ and\ \citenamefont {Wrachtrup}}]{baranov2017}%
  \BibitemOpen
  \bibfield  {author} {\bibinfo {author} {\bibfnamefont {P.~G.}\ \bibnamefont {Baranov}}, \bibinfo {author} {\bibfnamefont {H.~J.}\ \bibnamefont {Von~Bardeleben}}, \bibinfo {author} {\bibfnamefont {F.}~\bibnamefont {Jelezko}},\ and\ \bibinfo {author} {\bibfnamefont {J.}~\bibnamefont {Wrachtrup}},\ }\href@noop {} {\emph {\bibinfo {title} {Magnetic Resonance of Semiconductors and Their Nanostructures}}}\ (\bibinfo  {publisher} {Springer},\ \bibinfo {year} {2017})\BibitemShut {NoStop}%
\bibitem [{\citenamefont {Grachev}(1987)}]{grachev1987correct}%
  \BibitemOpen
  \bibfield  {author} {\bibinfo {author} {\bibfnamefont {V.~G.}\ \bibnamefont {Grachev}},\ }\bibfield  {title} {\bibinfo {title} {Correct expression for the generalized spin hamiltonian for a noncubic paramagnetic center},\ }\href@noop {} {\bibfield  {journal} {\bibinfo  {journal} {Soviet Journal of Experimental and Theoretical Physics}\ }\textbf {\bibinfo {volume} {65}},\ \bibinfo {pages} {1029} (\bibinfo {year} {1987})}\BibitemShut {NoStop}%
\bibitem [{\citenamefont {Stoll}\ and\ \citenamefont {Schweiger}(2006)}]{stoll2006easyspin}%
  \BibitemOpen
  \bibfield  {author} {\bibinfo {author} {\bibfnamefont {S.}~\bibnamefont {Stoll}}\ and\ \bibinfo {author} {\bibfnamefont {A.}~\bibnamefont {Schweiger}},\ }\bibfield  {title} {\bibinfo {title} {Easyspin, a comprehensive software package for spectral simulation and analysis in epr},\ }\href@noop {} {\bibfield  {journal} {\bibinfo  {journal} {Journal of magnetic resonance}\ }\textbf {\bibinfo {volume} {178}},\ \bibinfo {pages} {42} (\bibinfo {year} {2006})}\BibitemShut {NoStop}%
\bibitem [{\citenamefont {Babunts}\ \emph {et~al.}(2021)\citenamefont {Babunts}, \citenamefont {Anisimov}, \citenamefont {Breev}, \citenamefont {Gurin}, \citenamefont {Bundakova}, \citenamefont {Muzafarova}, \citenamefont {Mokhov},\ and\ \citenamefont {Baranov}}]{babunts2021fully}%
  \BibitemOpen
  \bibfield  {author} {\bibinfo {author} {\bibfnamefont {R.~A.}\ \bibnamefont {Babunts}}, \bibinfo {author} {\bibfnamefont {A.~N.}\ \bibnamefont {Anisimov}}, \bibinfo {author} {\bibfnamefont {I.~D.}\ \bibnamefont {Breev}}, \bibinfo {author} {\bibfnamefont {A.~S.}\ \bibnamefont {Gurin}}, \bibinfo {author} {\bibfnamefont {A.~P.}\ \bibnamefont {Bundakova}}, \bibinfo {author} {\bibfnamefont {M.~V.}\ \bibnamefont {Muzafarova}}, \bibinfo {author} {\bibfnamefont {E.}~\bibnamefont {Mokhov}},\ and\ \bibinfo {author} {\bibfnamefont {P.~G.}\ \bibnamefont {Baranov}},\ }\bibfield  {title} {\bibinfo {title} {Fully optical detection of hyperfine electron--nuclear interactions in spin centers in 6h-sic crystals with a modified $^{13}\text{C}$ isotope content},\ }\href@noop {} {\bibfield  {journal} {\bibinfo  {journal} {JETP Letters}\ }\textbf {\bibinfo {volume} {114}},\ \bibinfo {pages} {463} (\bibinfo {year} {2021})}\BibitemShut {NoStop}%
\bibitem [{\citenamefont {Soltamov}\ \emph {et~al.}(2021)\citenamefont {Soltamov}, \citenamefont {Yavkin}, \citenamefont {Mamin}, \citenamefont {Orlinskii}, \citenamefont {Breev}, \citenamefont {Bundakova}, \citenamefont {Babunts}, \citenamefont {Anisimov},\ and\ \citenamefont {Baranov}}]{soltamov2021electron}%
  \BibitemOpen
  \bibfield  {author} {\bibinfo {author} {\bibfnamefont {V.~A.}\ \bibnamefont {Soltamov}}, \bibinfo {author} {\bibfnamefont {B.~V.}\ \bibnamefont {Yavkin}}, \bibinfo {author} {\bibfnamefont {G.~V.}\ \bibnamefont {Mamin}}, \bibinfo {author} {\bibfnamefont {S.~B.}\ \bibnamefont {Orlinskii}}, \bibinfo {author} {\bibfnamefont {I.~D.}\ \bibnamefont {Breev}}, \bibinfo {author} {\bibfnamefont {A.~P.}\ \bibnamefont {Bundakova}}, \bibinfo {author} {\bibfnamefont {R.~A.}\ \bibnamefont {Babunts}}, \bibinfo {author} {\bibfnamefont {A.~N.}\ \bibnamefont {Anisimov}},\ and\ \bibinfo {author} {\bibfnamefont {P.~G.}\ \bibnamefont {Baranov}},\ }\bibfield  {title} {\bibinfo {title} {Electron nuclear interactions in spin-3/2 color centers in silicon carbide: A high-field pulse epr and endor study},\ }\href@noop {} {\bibfield  {journal} {\bibinfo  {journal} {Physical Review B}\ }\textbf {\bibinfo {volume} {104}},\ \bibinfo {pages} {125205} (\bibinfo {year} {2021})}\BibitemShut {NoStop}%
\bibitem [{\citenamefont {Schloss}\ \emph {et~al.}(2018)\citenamefont {Schloss}, \citenamefont {Barry}, \citenamefont {Turner},\ and\ \citenamefont {Walsworth}}]{schloss2018simultaneous}%
  \BibitemOpen
  \bibfield  {author} {\bibinfo {author} {\bibfnamefont {J.~M.}\ \bibnamefont {Schloss}}, \bibinfo {author} {\bibfnamefont {J.~F.}\ \bibnamefont {Barry}}, \bibinfo {author} {\bibfnamefont {M.~J.}\ \bibnamefont {Turner}},\ and\ \bibinfo {author} {\bibfnamefont {R.~L.}\ \bibnamefont {Walsworth}},\ }\bibfield  {title} {\bibinfo {title} {Simultaneous broadband vector magnetometry using solid-state spins},\ }\href@noop {} {\bibfield  {journal} {\bibinfo  {journal} {Physical Review Applied}\ }\textbf {\bibinfo {volume} {10}},\ \bibinfo {pages} {034044} (\bibinfo {year} {2018})}\BibitemShut {NoStop}%
\bibitem [{\citenamefont {Garsi}\ \emph {et~al.}(2024)\citenamefont {Garsi}, \citenamefont {St{\"o}hr}, \citenamefont {Denisenko}, \citenamefont {Shagieva}, \citenamefont {Trautmann}, \citenamefont {Vogl}, \citenamefont {Sene}, \citenamefont {Kaiser}, \citenamefont {Zappe}, \citenamefont {Reuter},\ and\ \citenamefont {Wrachtrup}}]{garsi2024three}%
  \BibitemOpen
  \bibfield  {author} {\bibinfo {author} {\bibfnamefont {M.}~\bibnamefont {Garsi}}, \bibinfo {author} {\bibfnamefont {R.}~\bibnamefont {St{\"o}hr}}, \bibinfo {author} {\bibfnamefont {A.}~\bibnamefont {Denisenko}}, \bibinfo {author} {\bibfnamefont {F.}~\bibnamefont {Shagieva}}, \bibinfo {author} {\bibfnamefont {N.}~\bibnamefont {Trautmann}}, \bibinfo {author} {\bibfnamefont {U.}~\bibnamefont {Vogl}}, \bibinfo {author} {\bibfnamefont {B.}~\bibnamefont {Sene}}, \bibinfo {author} {\bibfnamefont {F.}~\bibnamefont {Kaiser}}, \bibinfo {author} {\bibfnamefont {A.}~\bibnamefont {Zappe}}, \bibinfo {author} {\bibfnamefont {R.}~\bibnamefont {Reuter}},\ and\ \bibinfo {author} {\bibfnamefont {J.}~\bibnamefont {Wrachtrup}},\ }\bibfield  {title} {\bibinfo {title} {Three-dimensional imaging of integrated-circuit activity using quantum defects in diamond},\ }\href@noop {} {\bibfield  {journal} {\bibinfo  {journal} {Physical Review Applied}\ }\textbf {\bibinfo {volume} {21}},\ \bibinfo {pages} {014055} (\bibinfo {year}
  {2024})}\BibitemShut {NoStop}%
\bibitem [{\citenamefont {Widmann}\ \emph {et~al.}(2015)\citenamefont {Widmann}, \citenamefont {Lee}, \citenamefont {Rendler}, \citenamefont {Son}, \citenamefont {Fedder}, \citenamefont {Paik}, \citenamefont {Yang}, \citenamefont {Zhao}, \citenamefont {Yang}, \citenamefont {Booker}, \citenamefont {Denisenko}, \citenamefont {Jamali}, \citenamefont {Momenzadeh}, \citenamefont {Gerhardt}, \citenamefont {Ohshima}, \citenamefont {Gali}, \citenamefont {Janzén},\ and\ \citenamefont {Wrachtrup}}]{widmann2015coherent}%
  \BibitemOpen
  \bibfield  {author} {\bibinfo {author} {\bibfnamefont {M.}~\bibnamefont {Widmann}}, \bibinfo {author} {\bibfnamefont {S.-Y.}\ \bibnamefont {Lee}}, \bibinfo {author} {\bibfnamefont {T.}~\bibnamefont {Rendler}}, \bibinfo {author} {\bibfnamefont {N.~T.}\ \bibnamefont {Son}}, \bibinfo {author} {\bibfnamefont {H.}~\bibnamefont {Fedder}}, \bibinfo {author} {\bibfnamefont {S.}~\bibnamefont {Paik}}, \bibinfo {author} {\bibfnamefont {L.-P.}\ \bibnamefont {Yang}}, \bibinfo {author} {\bibfnamefont {N.}~\bibnamefont {Zhao}}, \bibinfo {author} {\bibfnamefont {S.}~\bibnamefont {Yang}}, \bibinfo {author} {\bibfnamefont {I.}~\bibnamefont {Booker}}, \bibinfo {author} {\bibfnamefont {A.}~\bibnamefont {Denisenko}}, \bibinfo {author} {\bibfnamefont {M.}~\bibnamefont {Jamali}}, \bibinfo {author} {\bibfnamefont {S.~A.}\ \bibnamefont {Momenzadeh}}, \bibinfo {author} {\bibfnamefont {I.}~\bibnamefont {Gerhardt}}, \bibinfo {author} {\bibfnamefont {T.}~\bibnamefont {Ohshima}}, \bibinfo {author} {\bibfnamefont {A.}~\bibnamefont
  {Gali}}, \bibinfo {author} {\bibfnamefont {E.}~\bibnamefont {Janzén}},\ and\ \bibinfo {author} {\bibfnamefont {J.}~\bibnamefont {Wrachtrup}},\ }\bibfield  {title} {\bibinfo {title} {Coherent control of single spins in silicon carbide at room temperature},\ }\href@noop {} {\bibfield  {journal} {\bibinfo  {journal} {Nature materials}\ }\textbf {\bibinfo {volume} {14}},\ \bibinfo {pages} {164} (\bibinfo {year} {2015})}\BibitemShut {NoStop}%
\end{thebibliography}%

\newpage

\section*{Figure captions}

\begin{figure*}
\includegraphics[width=0.7\linewidth]{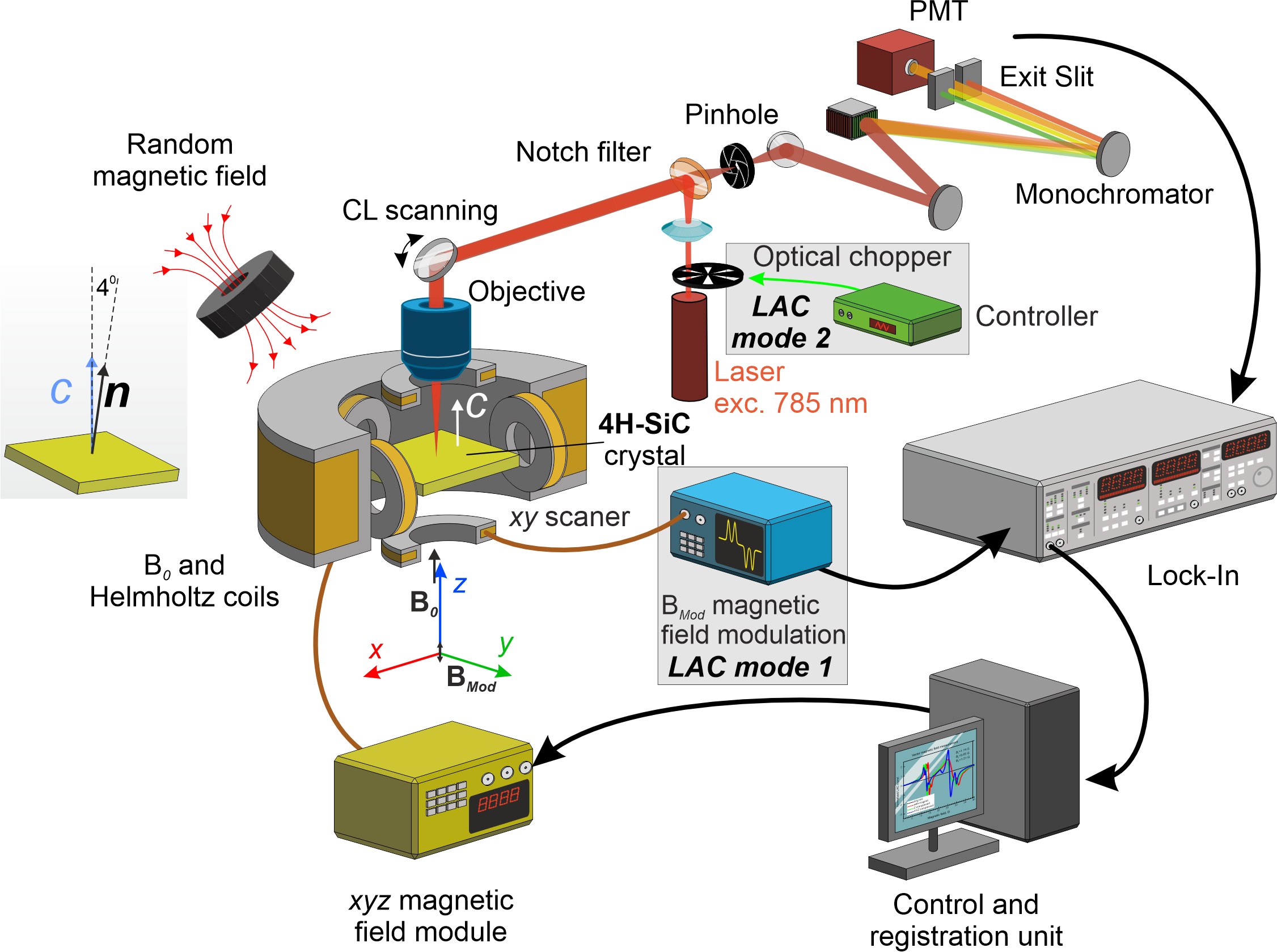} 
\caption{\label{fig:Block diagram} }
\end{figure*}

Figure~\ref{fig:Block diagram}. Block diagram of the all-optical vector magnetometer. 
There are two modes for level anti-crossing (LAC) signal acquisition. In mode 1 the 
quasi-stationary magnetic field ${B}_{0}$ is modulated at low frequency (LF).
In mode 2 the intensity of the exciting laser radiation is modulated at low frequency. 
For both options, the signal is recorded at low frequency using a lock-in detector. 
4H-SiC wafer with spin-$\frac{3}{2}$ centers is used as the SiC sensor for measurements, 
crystal $c$~axis perpendicular to the wafer plane. The inset shows a structure in which 
a 4H-SiC wafer is used as the SiC sensor for measurements, in which the $c$~axis is 
deviated from the surface normal by 4 degrees (this configuration 
is often found in SiC device structures). The optical magnetometer includes an active 
material in the form of a silicon carbide crystal wafer of a hexagonal or orthorhombic 
polytype containing spin-$\frac{3}{2}$ centers and placed in an electromagnet that scans 
the a quasi-stationary magnetic field ${B}_{0}$ along the 
$z$~axis, which is combined with the electromagnet in the form of 
Helmholtz coils, creating a field along the ${x, y, z}$~axes. These 
coils are designed to create an external magnetic field when calibrating the magnetometer, 
as well as to create a compensating magnetic field during measurements. A modulated 
magnetic field ${B}_{Mod}$ can be applied to the coil for 
$z$~axis. As the low-frequency driving current in the Helmholtz coils 
is directly proportional to the magnetic field it generates, current value will serve 
as an indirect measure of the magnetic field being cancelled in each dimension.

\begin{figure}
\includegraphics{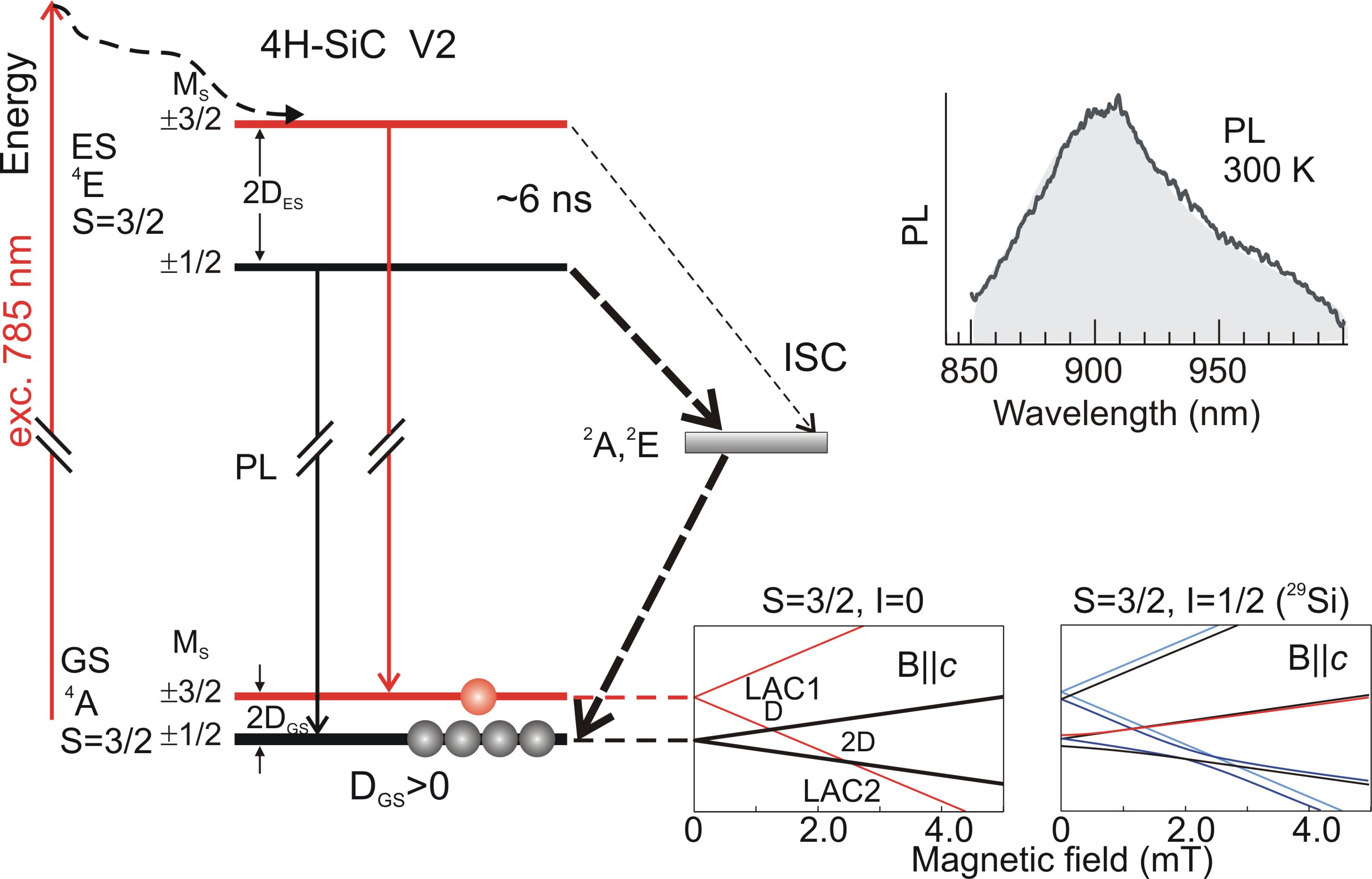}
\caption{\label{fig:Diagram of the energy levels} }
\end{figure}

Figure~\ref{fig:Diagram of the energy levels}. Diagram of the energy levels of the V2 spin center in 4H-SiC,
having $S ={3}/{2}$ both in the ground state (GS) and in the excited state (ES). The cycle of optically induced 
alignment of spin level populations in the ground state is shown: excitation with spin conservation, emission 
with spin conservation, non-radiative recombination through an intermediate metastable state 
(inter-system crossing - ISC) without spin conservation. As a result, the lower levels with $M_{S}= \pm 1/2$
are populated, and the levels $M_{S}= \pm 3/2$ are emptied. Two diagrams of energy levels in a magnetic field 
are shown for even silicon isotopes ($I=0$) and taking into account hyperfine interactions with one nucleus 
of the $^{29}$Si isotope ($I=1/2$).

\begin{figure}
\includegraphics{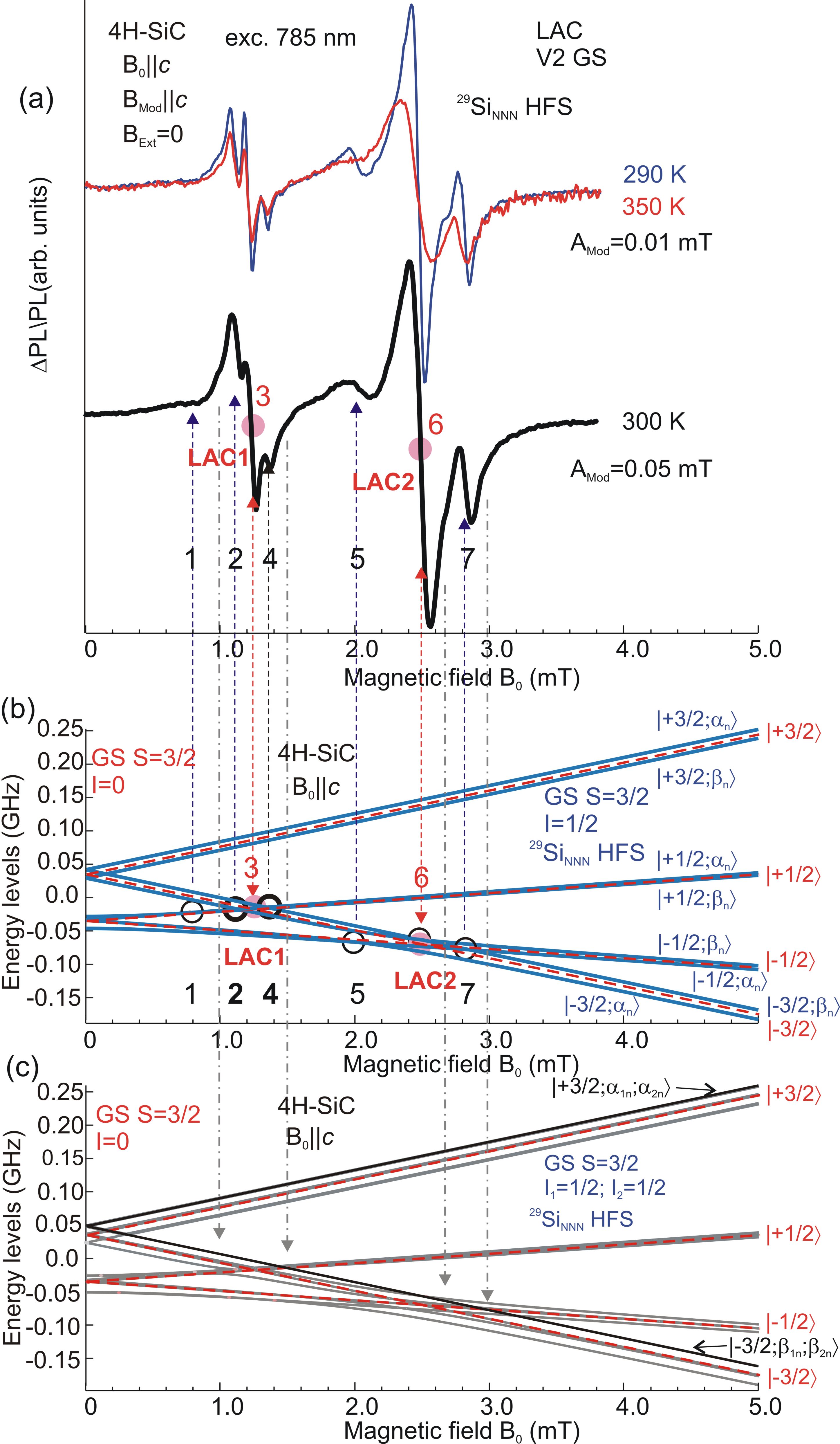}
\caption{\label{fig:Dependence PL}}
\end{figure}

Figure~\ref{fig:Dependence PL}. (a) Dependence of the photoluminescence (PL) intensity of axial spin 
centers V2, recorded in a 4H-SiC crystal with a natural content of isotopes. Low-frequency modulation
of the magnetic field ${B}_{Mod}$ is used, the direction of which is coaxial with the 
direction of the quasi-stationary magnetic field ${B}_{0}$. The two upper spectra were recorded 
at a low temperature of 290~K (blue line) and a high temperature of 350~K (red line) using a small 
magnetic field modulation amplitude, less than 0.01~mT. Below is the LAC spectrum recorded at 
a temperature of 300~K (black line) and a large magnetic field modulation amplitude, $\sim$0.05~mT. 
(b) Energy levels calculated for the system $S=3/2$: solid lines for $I=1/2$; dotted lines for $I=0$. 
The filled pink circles show the level anti-crossings LAC1 and LAC2 for the ground state in the absence
of hyperfine interaction. 
Open black circles show LACs in the presence of one silicon atom with a nuclear magnetic moment 
of the $^{29}$Si isotope in the second coordination sphere. (c) Energy levels for a system with $S=3/2$
and hyperfine interaction with two $^{29}$Si nuclei, $I_{1}=1/2$ and $I_{2}=1/2$, which with 
a much lower probability can be included in the structure of the spin-$\frac{3}{2}$ center.

\begin{figure}
\includegraphics{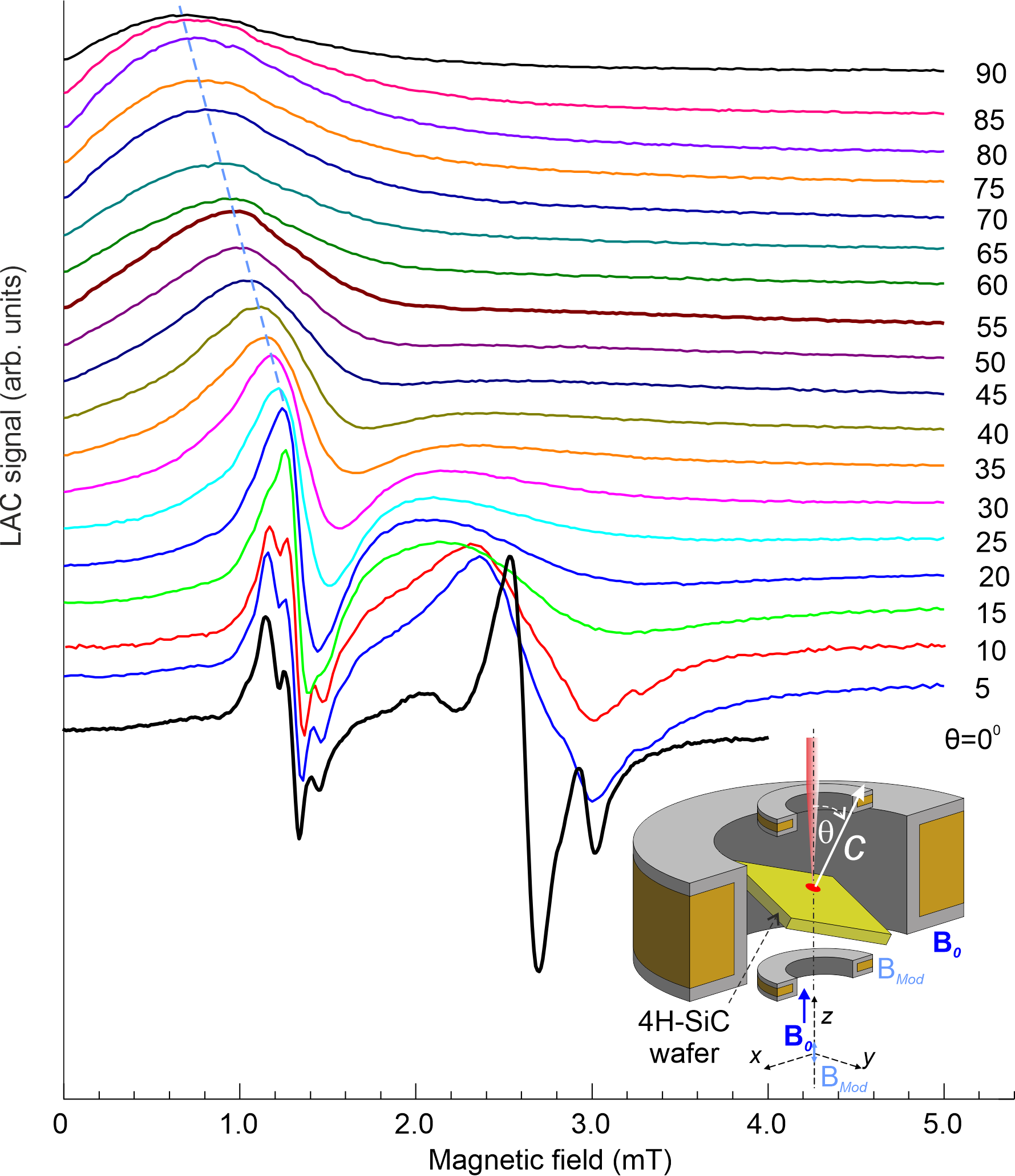}
\caption{\label{fig:Orientation dependences}}
\end{figure}

Figure~\ref{fig:Orientation dependences}. Orientation dependences of the LAC spectra when the 
crystal rotates in a quasi-stationary magnetic field ${B}_{0}$, as shown in the inset. 
The direction of the low-frequency modulation magnetic field ${B}_{Mod}$ coincides 
with ${B}_{0}$.

\begin{figure}
\includegraphics{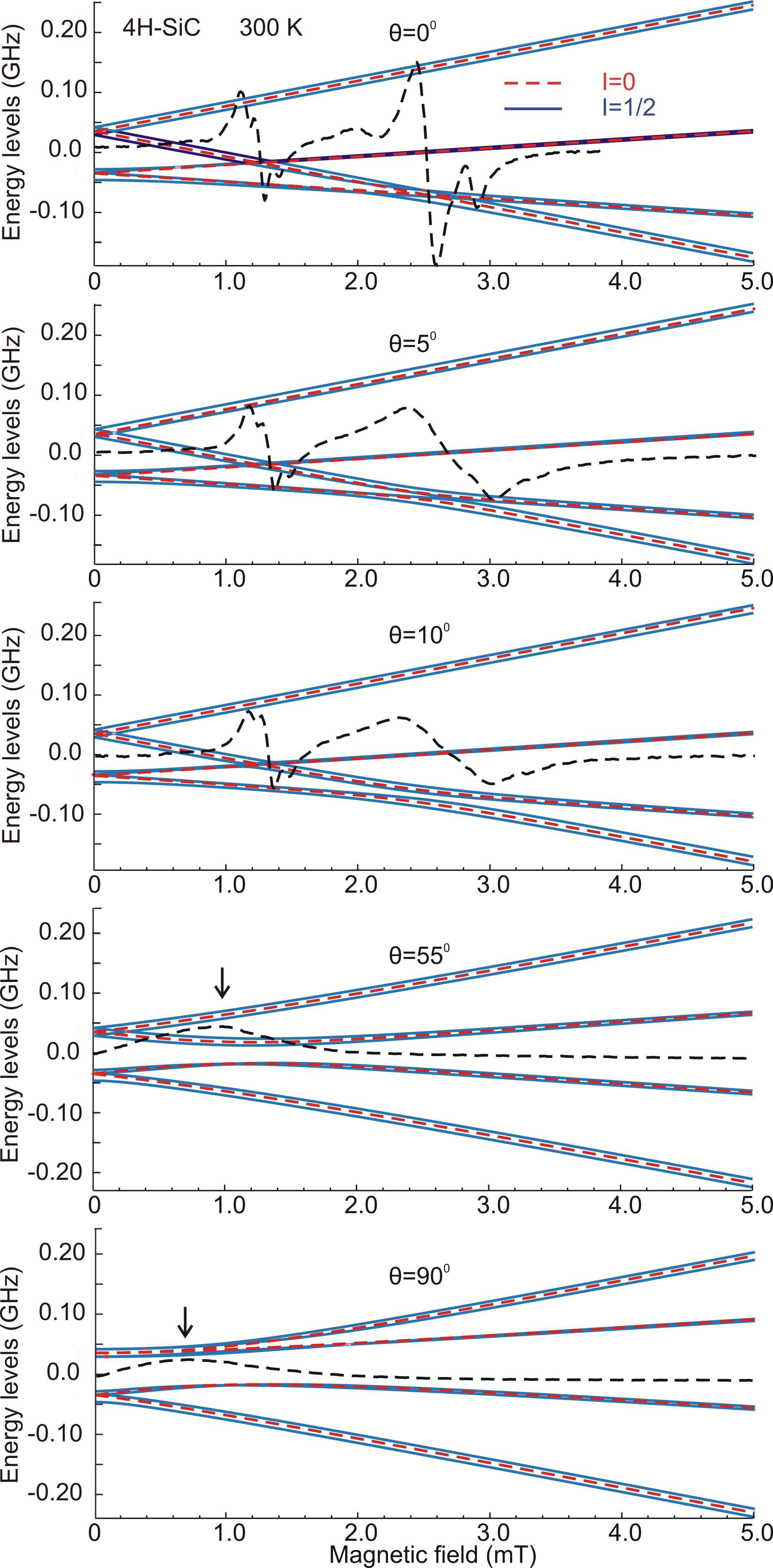}
\caption{\label{fig:energy levels calc} }
\end{figure}

Figure~\ref{fig:energy levels calc}. The energy levels for spin-$\frac{3}{2}$ center 
in a 4H-SiC calculated for different crystal orientations in a magnetic field. Solid lines 
correspond to the structure of the spin center, in which there is hyperfine interaction 
with one $\frac{3}{2}$ nucleus (Si$_{NNN}$), $I=1/2$; dotted lines correspond to a structure 
in which there is no hyperfine interaction, since there are no $^{29}$Si nuclei in the second 
coordination sphere of twelve silicon atoms, $I=0$. The dotted black lines show the corresponding 
LAC spectra taken from Fig.~\ref{fig:Orientation dependences}.

\begin{figure}
\includegraphics{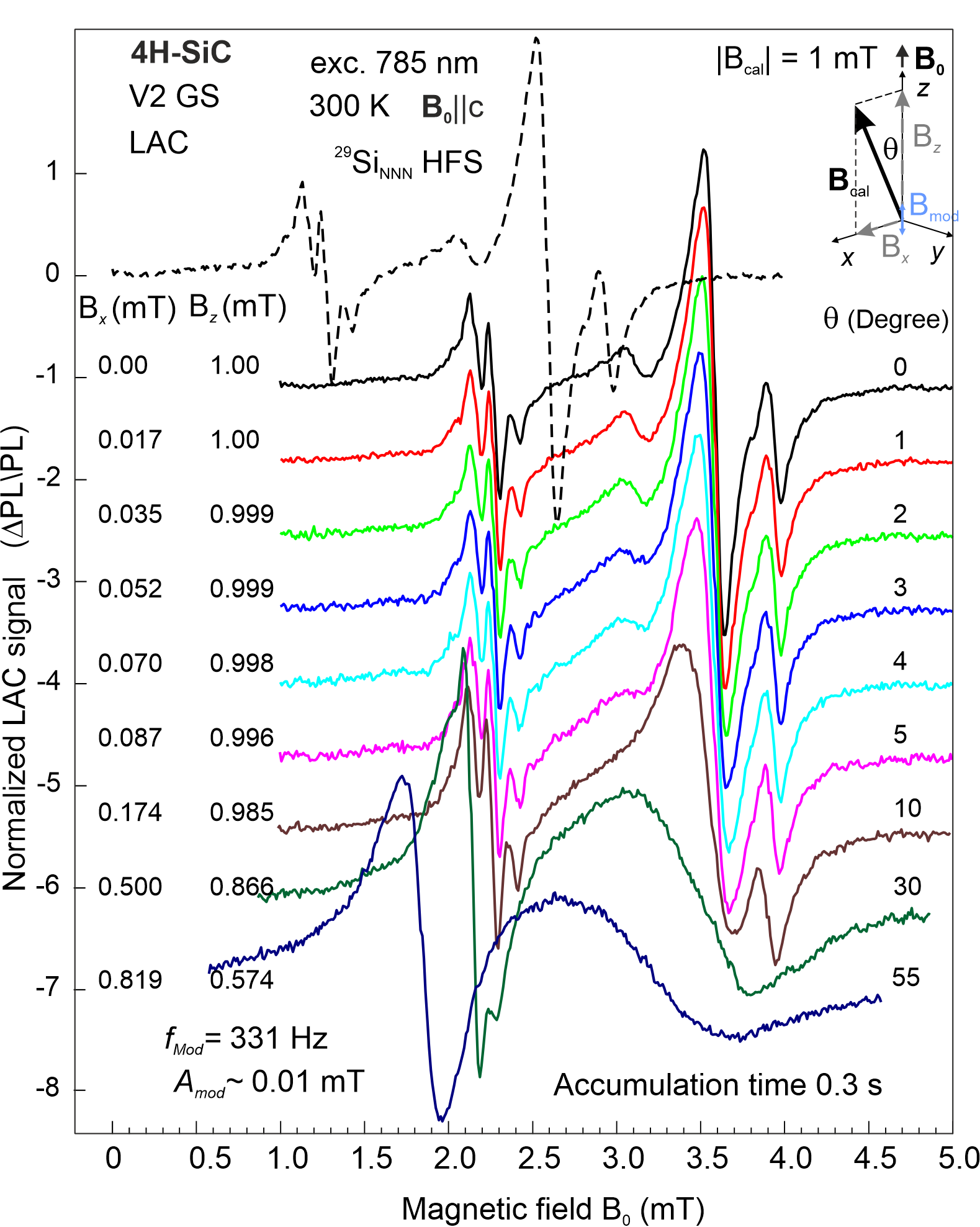}
\caption{\label{fig:principles of calibrating} }
\end{figure}

Figure~\ref{fig:principles of calibrating}. Demonstration of the principles of calibrating 
a vector magnetometer using a system of magnetic Helmholtz coils to create a given magnetic 
field in three directions  ${x, y, z}$. For ease of measurement, the calibrated magnetic
field will be applied in the  $z$ and  ${x}$ directions, resulting in a change 
in the polar angle $\theta$. Since the measurements are based on axial spin centers whose axis 
coincides with the $c$~axis, the measurement results depend only on the ratio of the longitudinal 
and perpendicular components of the calibration magnetic field, that is, they depend on the polar
angle $\theta$ and do not depend on the azimuthal angle $\varphi$. The dotted line shows the 
reference LAC spectrum recorded in a zero external magnetic field in the orientation of a 
quasi-stationary magnetic field ${B}_{0}$ and a modulation magnetic field ${B}_{Mod}$
along the $c$~axis of 4H-SiC crystal used as a magnetic sensor.

\begin{figure}
\includegraphics[width=0.6\linewidth]{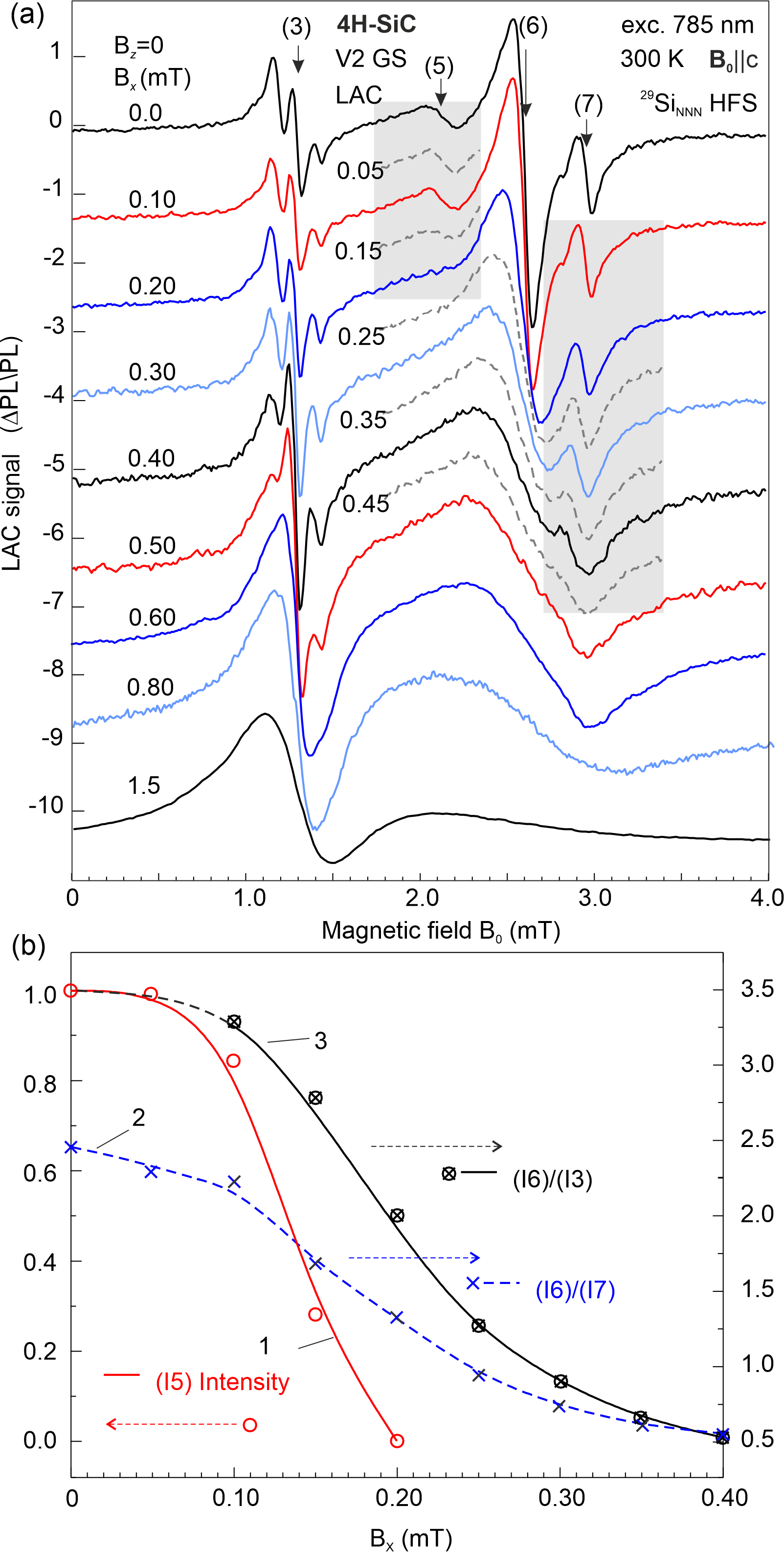}
\caption{\label{fig:Calibrating magnetometer} }
\end{figure}

Figure~\ref{fig:Calibrating magnetometer}. (a) Calibrating a vector magnetometer when the 
$z$-component of the magnetic field ${B}_{z}$ is reset to zero by shifting the LAC1 
point to the standard value $D$ (1.25 mT). Next, only the perpendicular component changes. For ease
of measurement, the calibrated perpendicular magnetic field will be applied in ${x}$ directions,
${B}_{x}$ mT (${B}_{z}=0$). (b) Changes in the intensities $I$ of selected lines in the
LAC spectra (3, 5, 6, 7 in the top of Fig.~\ref{fig:Calibrating magnetometer}(a)), characterizing changes
in the photoluminescence intensity of spin centers with a ground quadruplet spin-$\frac{3}{2}$ state,
depending on perpendicular magnetic field ${B}_{x}$ (mT), which serve as criteria for the 
preliminary assessment of external measured magnetic fields and the selection of the values of 
compensating magnetic fields ${B}_{Comp}$ for zeroing the measured external magnetic field.

\begin{figure}
\includegraphics{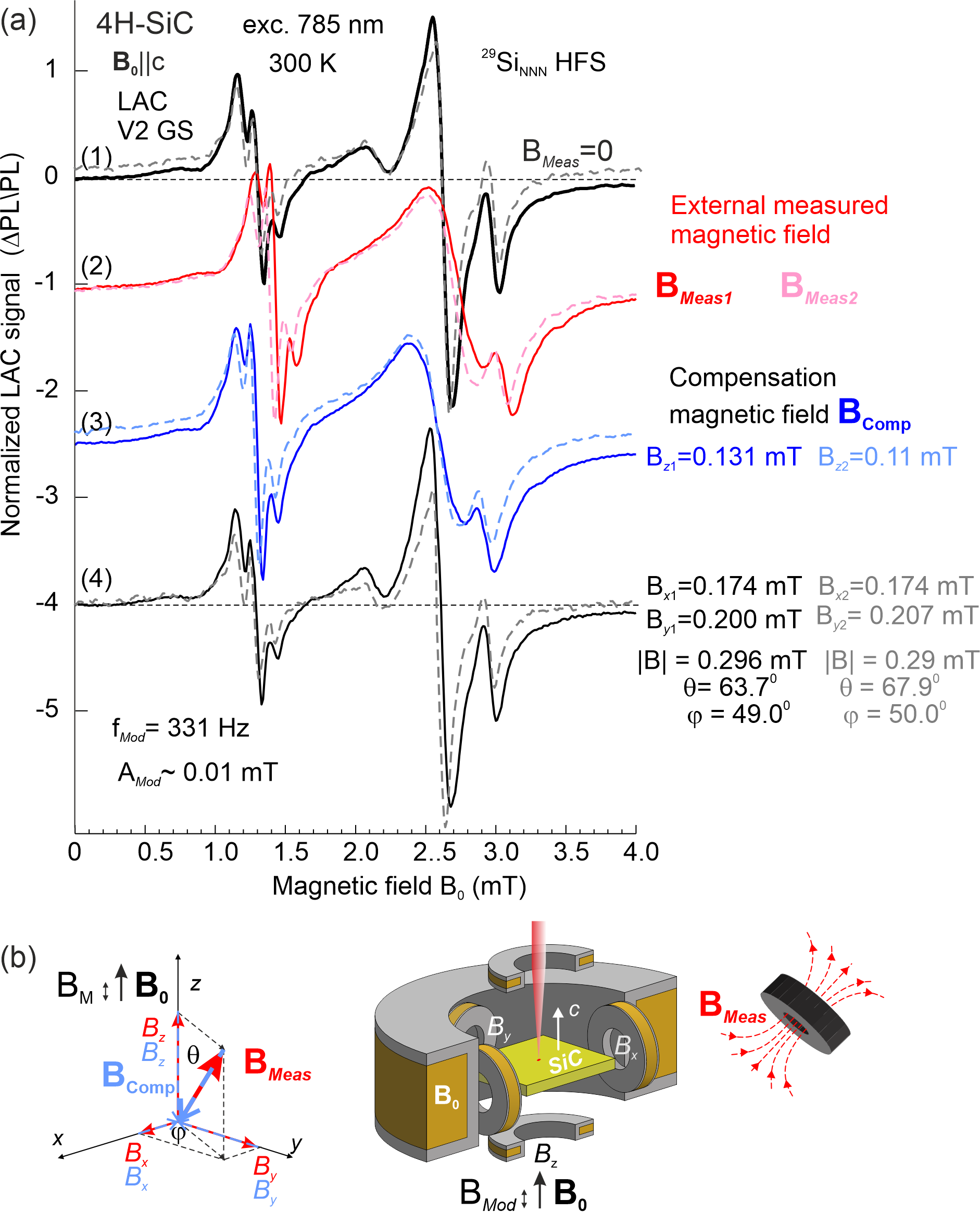}
\caption{\label{fig:Demon princ} }
\end{figure}

Figure~\ref{fig:Demon princ}. Demonstration of the principles of all-optical quantum 
vector magnetometry. Two experiments ($\#$1 and $\#$2) are presented to compensate for the external random 
created magnetic field by two positions of a permanent magnet. (a) Top shows the spectrum of LAC (1) 
recorded in the absence of an external magnetic field (solid lines for all spectra correspond to 
experiment $\#$1, dotted lines to experiment $\#$2), also shown in Fig.~\ref{fig:Dependence PL}, where 
the energy levels leading to mentioned LAC were demonstrated. The following spectra (2) were recorded 
with the application of an arbitrary external measured magnetic fields ${B}_{Meas_{1}}$ and
${B}_{Meas_{2}}$, conventionally shown in (b). The two lower spectra correspond to the sequential 
application of compensating magnetic fields in the form of the ${B}_{z_{1}}$ and ${B}_{z_{2}}$
component (3), shifting the signals of the first LAC1 until they coincide with the initial signal (1),
and then the ${B}_{x_{1}}$ and ${B}_{y_{1}}$ (${B}_{x_{2}}$ and ${B}_{y_{2}}$) 
components (4) are applied until a spectrum appears that almost coincide with the spectra (1), shown
with a dotted line. (b) The principle of compensating a random magnetic field ${B}_{Meas}$ 
(red arrows) by applying a compensation field ${B}_{Comp}$ (blue arrows) is shown conventionally.

\begin{figure}
\includegraphics{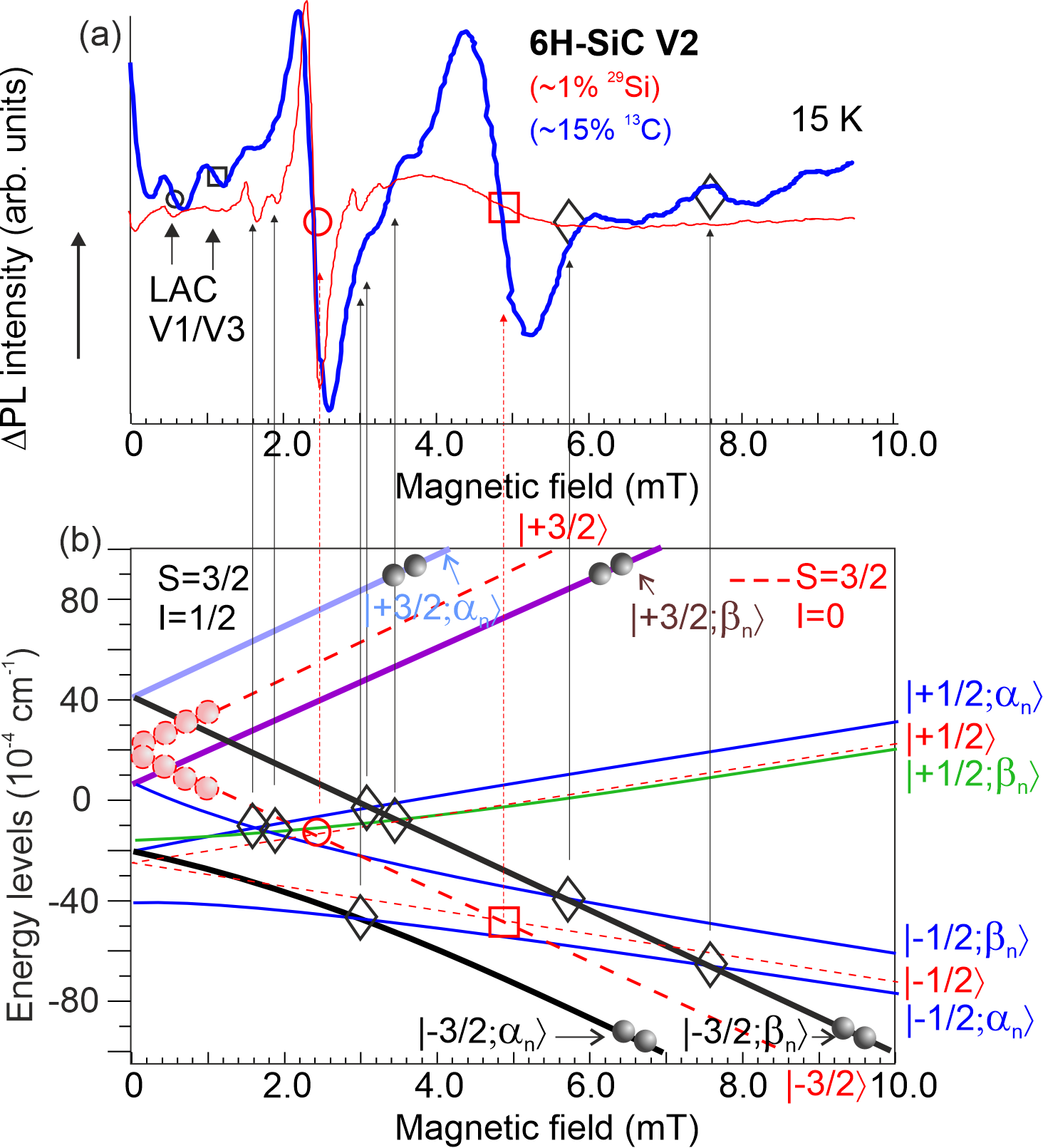}
\caption{\label{fig:LAC recorded} }
\end{figure}

Figure~\ref{fig:LAC recorded}. (a) Level anticrossing spectra recorded in the ground state for V2 centers 
in a 6H-SiC crystal, in the $B\parallel{c}$ orientation, excitation 808~nm, grown with a depleted $^{29}$Si
isotope content, $\sim1\%$, natural $^{13}$C isotope content, 1.1$\%$ (red color) and in a 6H-SiC crystal
grown with an enriched content of the $^{13}$C isotope up to $\sim15\%$, the natural content of the 
$^{29}$Si isotope is $4.7\%$ (blue color). Open circles and squares indicate anticrossings of LAC1 and LAC2
levels, respectively; circles and squares of reduced size show signals of level anti-crossing for V1/V3 centers.
(b) Scheme of energy levels in the magnetic field of V2 centers taking into account hyperfine interactions with
the one nuslear $^{13}$C isotope ($I = 1/2$), anti-crossing signals are marked by open diamonds. The dotted line in red 
shows energy levels in a magnetic field without taking into account hyperfine interactions.
It can be seen that a decrease in the content of the $^{29}$Si isotope, which has a nuclear magnetic moment, 
significantly narrows the line. It is also important to emphasize the gigantic increase in line widths in 
a crystal enriched with the $^{13}$C isotope. Thus, it is possible to reduce the width of the LAC line and 
thereby increase the accuracy of measuring magnetic fields by changing the isotopic composition of SiC, 
in particular, reducing the content of 
the $^{13}$C isotope.

\begin{figure}
\includegraphics{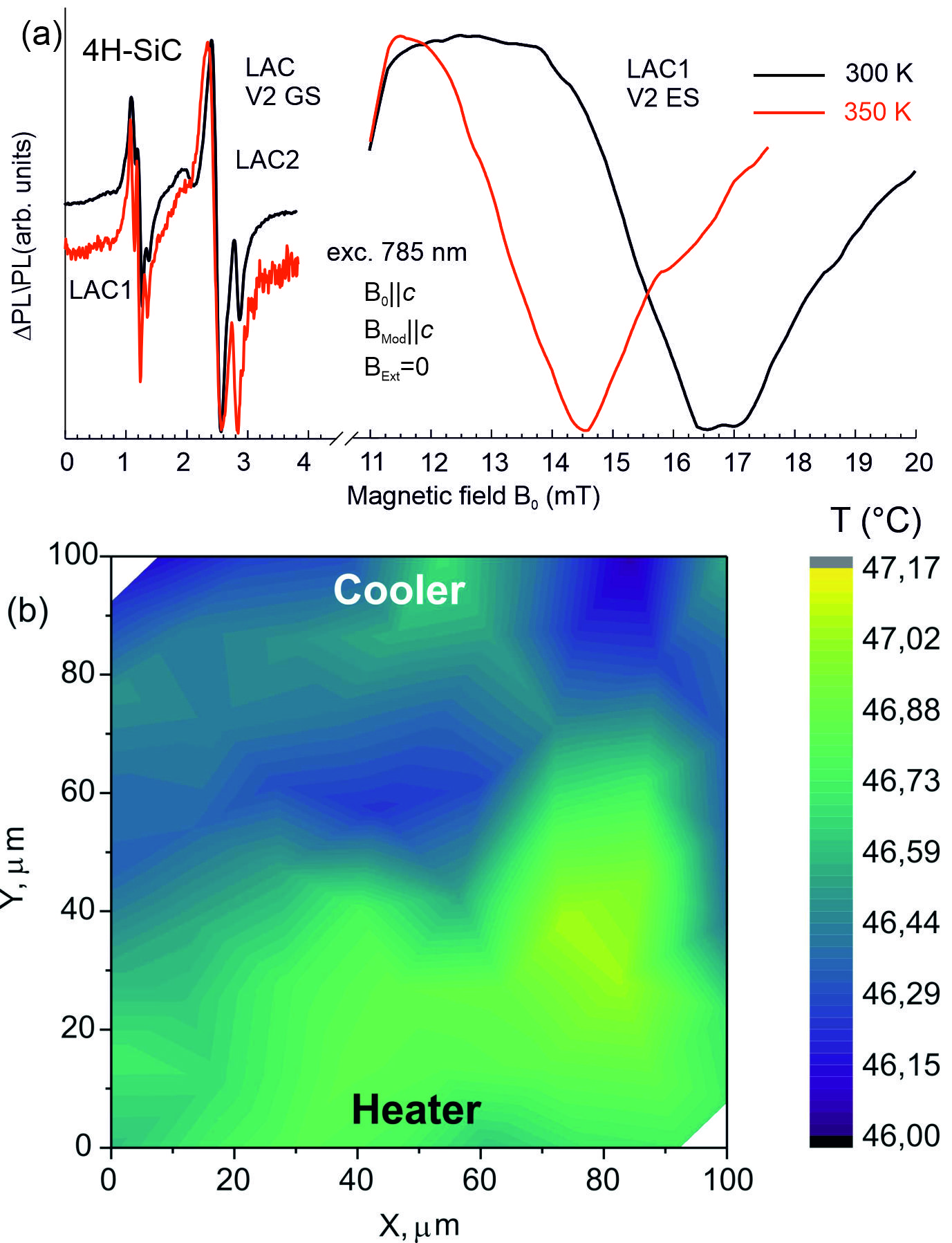}
\caption{\label{fig:Temperature} }
\end{figure}

Figure~\ref{fig:Temperature}. (a) The effect of temperature on the position of LAC in the ground 
and excited states in a 4H-SiC crystal. Shown is a ground state reference spectrum recorded at two 
temperatures, room temperature 300~K and elevated temperature 350~K, and an excited state LAC 
spectrum recorded at the same temperatures.
(b) The temperature color map obtained by converting the position of LAC excited states 
into a temperature value from 
a sample area of $100\times100~\mu\text{m}$ with 
a step of $5~\mu\text{m}$.

\newpage

\end{document}